\documentclass[letterpaper,twocolumn,10pt,colorlinks=true,linkcolor=black]{article}
\usepackage{xurl}
\usepackage{usenix}

\def\showcomments{0}
\def\techreport{1} 

\ifnum\techreport=0
\fi

\usepackage{url}

\usepackage{graphicx}
\usepackage{diagbox}

\usepackage{xspace}
\usepackage{xcolor} 
\usepackage{tabularx}
\usepackage{subfig}
\usepackage{enumitem}
\usepackage[T1]{fontenc}
\usepackage{hyperref}
\usepackage{xurl}
\hypersetup{breaklinks=true}
\usepackage{amsmath}
\usepackage{tikz}
\usepackage{soul}
\usepackage{multirow}
\usepackage{booktabs}
\usepackage{csquotes} 
\usepackage[normalem]{ulem} 
\usepackage[capitalize,noabbrev]{cleveref}

\usepackage{listings}
\usepackage{siunitx} 
\usepackage{pdflscape}
\usepackage{pgfplotstable}
\usepackage{pgfplots}
\usepackage{pgfplotstable}
\usepgfplotslibrary{colorbrewer}
\usetikzlibrary{patterns}

\usepackage{etoolbox}
\preto\subequations{\ifhmode\unskip\fi}

\newcommand{\bheading}[1]{{\vspace{4pt}\noindent\textbf{#1}}}

\ifnum\showcomments=1
    \newcommand{\gennote}[4][blue]{\textcolor{#1}{$\rule{8pt}{8pt}_{\textsf{\scshape\bfseries #2}}$ \textcolor{gray}{\emph{\sout{#3}}}#4}}
    \newcommand{\liangw}[1]{{\textcolor{red}{[LW: #1]}}}
    \newcommand{\david}[1]{{\textcolor{red}{[david: #1]}}}
    \newcommand{\henry}[1]{\gennote[green!40!black]{henry}{}{[#1]}}
    \newcommand{\minorrevisions}[1]{\gennote[green!40!black]{minor-revisions}{}{[#1]}}
    \newcommand{\henrynew}[1]{\gennote[green!40!black]{henry-new}{}{[#1]}}
    \newcommand{\joel}[1]{\gennote[orange]{joel}{}{[#1]}}
    \newcommand{\note}[1]{\gennote{note}{}{[#1]}}
    \newcommand{\todo}[1]{\gennote[red]{todo}{}{[#1]}}
    \newcommand{\adrian}[1]{\gennote[blue]{adrian}{}{[#1]}}
    \newcommand{\prateek}[1]{\gennote[red]{prateek}{}{[#1]}}
    \newcommand{\yixin}[1]{\gennote[purple]{yixin}{}{[#1]}}
    \newcommand{\grace}[1]{\gennote[magenta]{grace}{}{[#1]}}
    \newcommand{\kwon}[1]{{\textcolor{brown}{[JK: #1]}}}

\else
    \newcommand{\gennote}[4][blue]{}
    \newcommand{\liangw}[1]{}
    \newcommand{\david}[1]{}
    \newcommand{\henry}[1]{}
    \newcommand{\minorrevisions}[1]{}
    \newcommand{\henrynew}[1]{}
    \newcommand{\joel}[1]{}
    \newcommand{\note}[1]{}
    \newcommand{\todo}[1]{}
    \newcommand{\adrian}[1]{}
    \newcommand{\prateek}[1]{}
    \newcommand{\yixin}[1]{}
    \newcommand{\grace}[1]{}
    \newcommand{\kwon}[1]{}

\fi

\setlist[itemize]{leftmargin=1em, itemsep=-0.1em, itemindent=0em}
\setlist[enumerate]{leftmargin=1.2em, itemindent=0em, itemsep=-0.1em}
\setlist[description]{leftmargin=1em}

\renewcommand{\paragraph}[1]{\noindent\textbf{#1.}}

\newcounter{lineno}

\PassOptionsToPackage{hyphens}{url}

\usepackage[english]{babel}

\begin{document}

\title{\Large \bf Creating a Secure Underlay for the Internet}


\makeatletter
\def\and{%
  \end{tabular}%
  \hskip .1em \@plus.17fil\relax
  \begin{tabular}[t]{c}}
\makeatother

\author{
{\rm Henry Birge-Lee}\\
Princeton University
\and
{\rm Joel Wanner}\\
ETH Zürich
\and
{\rm Grace Cimaszewski}\\
Princeton University
\and
{\rm Jonghoon Kwon}\\
ETH Zürich
\and
{\rm Liang Wang}\\
Princeton University
\and
{\rm François Wirz}\\
ETH Zürich
\and
{\rm Prateek Mittal}\\
Princeton University
\and
{\rm Adrian Perrig}\\
ETH Zürich
\and
{\rm Yixin Sun}\\
University of Virginia

} 

\maketitle




%

\maketitle

\begin{abstract}
Adversaries can exploit inter-domain routing vulnerabilities to intercept communication and compromise the security of critical Internet applications.
Meanwhile the deployment of secure routing solutions such as Border Gateway Protocol Security (BGPsec) and Scalability, Control and Isolation On Next-generation networks (SCION) are still limited.
How can we leverage emerging secure routing backbones and extend their security properties to the broader Internet?
    
We design and deploy an architecture to bootstrap secure routing. 
Our key insight is to abstract the secure routing backbone as a virtual Autonomous System (AS), called Secure Backbone AS (SBAS). While SBAS appears as one AS to the Internet, it is a federated network where routes are exchanged between participants using a secure backbone. SBAS makes BGP announcements for its customers' IP prefixes at multiple locations (referred to as Points of Presence or PoPs) allowing traffic from non-participating hosts to be routed to a nearby SBAS PoP (where it is then routed over the secure backbone to the true prefix owner). In this manner, we are the first to integrate a federated secure non-BGP routing backbone with the BGP-speaking Internet.

We present a real-world deployment of our architecture that uses SCIONLab to emulate the secure backbone and the PEERING framework to make BGP announcements to the Internet. 
A combination of real-world attacks and Internet-scale simulations shows that SBAS substantially reduces the threat of routing attacks. Finally, we survey network operators to better understand optimal governance and incentive models.
\end{abstract}

\section{Introduction}
\label{sec:intro}

The de facto inter-domain routing protocol, the Border Gateway Protocol (BGP), is infamously insecure. Adversaries can exploit vulnerabilities in BGP to advertise bogus routes and hijack or intercept communications towards a victim~\cite{gavrichenkov2015breaking,nordstrom2004beware,birgelee2019sico,ars20183ve,goldberg2010secure}.
The initial secure routing efforts have focused on achieving \emph{origin validation}, i.e., validating the owner---the origin Autonomous Systems (AS)---of an IP prefix, in order to prevent prefix hijacking attacks. A standardized mechanism is the Resource Public Key Infrastructure (RPKI), which generates records that bind an IP prefix to the origin AS~\cite{rfc6810}. 
However, origin validation is insufficient for preventing more sophisticated interception attacks that manipulate routing paths~\cite{lychev2013bgp}. Recent works demonstrate the severity of interception attacks, including surveillance and compromising critical internet applications~\cite{arnbak2014loopholes,birge2018bamboozling,sun2015raptor,apostolaki2017bitcoin}.
Other proposals achieve path security in the routing backbone by authenticating the entire path information. Border Gateway Protocol Security (BGPsec) augments BGP by cryptographically signing and validating BGP paths~\cite{austein2017rfc}. However, BGPSec requires significant changes to the existing routing infrastructure and has yet to see production deployment. Private backbones and clean-slate Internet architectures such as Scalability, Control and Isolation On Next-generation networks (SCION)~\cite{Perrig2017} have also been proposed. While they are deployed in production networks, they cannot yet be used pervasively. 



In this context, can we design a system usable in today's Internet for improving routing security by leveraging an emerging secure routing backbone, such as SCION?
Toward this goal, we propose the Secure Backbone AS (SBAS), a novel
federated backbone infrastructure. 
SBAS abstracts away the secure backbone as a virtual AS that interacts
with traditional ASes through conventional BGP.
Within SBAS, the secure backbone allows for routing between participating customers and is immune to BGP attacks.
SBAS ensures that announcements received from participating customers are indeed authorized using existing techniques such as RPKI, and SBAS customers will prioritize routes received from SBAS.
Although the customers' connection to SBAS may cross a short
tunnel, we demonstrate that even in this scenario SBAS offers significant security benefits.

More specifically, a participating customer (which can be either an AS or a single end host) connects to SBAS via a secure tunnel---a Virtual Private Network (VPN)---if it is not a direct neighbor.
Participating customers can connect to SBAS via one or multiple distributed Points of Presence (PoPs).
A customer can bring its own prefix to announce through SBAS (e.g., an AS with its own address space), or use IP prefixes assigned by SBAS (e.g., individual clients and servers without control over their address space).
Within SBAS, customer address space is distributed via an SBAS-internal iBGP mesh between the PoPs allowing PoPs to announce customer prefixes to traditional BGP neighbors (providing connectivity and improved security for non-participating hosts).
We emphasize that SBAS is compatible with conventional BGP. This enables SBAS to route traffic between SBAS customers and non-participating ASes,providing security benefits even when one communication endpoint does not participate in SBAS.

We have implemented and deployed SBAS on real networks using SCIONLab~\cite{kwon2020scionlab} to emulate a secure SCION backbone, the PEERING framework~\cite{schlinker2019peering} to send/receive BGP announcements from non-participating ASes, and the WireGuard VPN to establish secure tunnels with SBAS customers.
Our implementation minimizes the need for new software and composes existing networking components to implement SBAS routing. 
Our key evaluation results are as follows:
\vspace{-2mm}
\begin{itemize}
  \item In our proof-of-concept deployment using SCIONLab and PEERING, we perform BGP attacks on IP prefixes of SBAS customers and of the VPN endpoints (in an ethical manner). SBAS successfully protects all customer-to-customer communication from our attacks and significantly improves the resilience of communication between SBAS customers and non-participating hosts. 
  \item Our Internet topology simulations further confirm that SBAS improves resilience to routing attacks on communication with \emph{non-participating} Internet hosts.
  Using a SBAS deployment with just six PoPs improves resilience by 61.8\%.
  Furthermore, SBAS integrates well with the existing effort on Route Origin Validation (ROV): if the broader Internet enforces ROV, 98.5\% of adversaries are  topologically incapable of hijacking SBAS-announced routes.
  \item Our proof-of-concept deployment only incurs an 11\% latency overhead on average (compared to the Internet), which decreases as more SBAS PoPs are deployed (as participants are closer to their nearest PoP). 
\end{itemize}
\vspace{-2mm}

\noindent An important benefit of the SBAS architecture is its compatibility with diverse secure (possibly non-BGP) backbone approaches~\cite{austein2017rfc,Perrig2017}. 
Moreover, given its strong security benefits with only a small latency overhead, SBAS represents a promising new abstraction for securing inter-domain routing that can provide much needed momentum and accelerate real-world adoption of secure backbones. To better understand the path towards a production deployment of SBAS, we surveyed network operators on appropriate incentives and governance structures, and found a potential community of early adopters as well as viable governance models.
Given the promising experimental and simulation results of our proof-of-concept SBAS implementation, we dare the community of network operators to realize SBAS in a production environment.

\section{Overview of Interdomain Routing Security}
\label{sec:background}

\textbf{BGP and BGP attacks.}
BGP is the inter-domain routing protocol today.
However, BGP lacks authentication of routing information, which allows for BGP attacks where an adversary maliciously sends BGP updates to hijack or intercept traffic to a victim AS~\cite{sun-cacm21}. Research shows that BGP attacks can have devastating consequences on critical Internet applications, including those that use
cryptographic security mechanisms~\cite{birge2018bamboozling,
  sun2015raptor, apostolaki2017bitcoin}. Also, BGP attacks are
routinely seen in the wild, impacting availability of Internet services and generating millions in revenue for
miscreants~\cite{ars20183ve, bgpmon2014canadian, myetherwallet}.

In \emph{equally-specific} BGP attacks, the adversary makes a malicious BGP
announcement for a victim's prefix that has the same prefix length
as the victim's prefix. 
Consequently, traffic may reach either the adversary or the victim depending on the routing polices. 
In \emph{more-specific} BGP attacks, the adversary announces a longer prefix than the victim's prefix. 
Because forwarding is based on longest prefix match,
traffic destined for the more specific prefix will be routed to the adversary.
This enables an adversary from almost any location in the Internet to attract a significant amount of network traffic destined to the victim. 
While more-specific BGP attacks are highly effective, they are not always viable given that most routers filter BGP announcements for prefixes longer than $/24$~\cite{ripe_prefix, cidr_report} (thus protecting $/24$ prefixes from more-specific attacks), and RPKI can also be used to filter malicious more-specific prefix announcements~\cite{rfc6810}.

The adversary may drop/respond to the traffic (\emph{hijack} attack) or 
forward the traffic back to the victim (\emph{interception} attack) via tunneling or existing BGP paths by deliberately shaping the malicious BGP announcements~\cite{birgelee2019sico, pilosov2008stealing}. 
Interception attacks are more sophisticated but also stealthier because the victim may see little to no difference (other than potentially increased latency) in its data plane traffic.

\textbf{Current Secure Routing: RPKI and route filtering.}
RPKI mitigates BGP attacks by providing a cryptographically secure database of IP address ownership that can be used to filter out bogus announcements~\cite{rpki}. In RPKI, each AS has a public--private key pair that is used to sign IP Route Origin Authorizations (ROAs) that associate IP prefixes with the ASes of their authorized origin ASes. The ASes compile the databases into a set of route filters  which block announcements that do not contain a valid $\langle$IP address, origin$\rangle$ pair. 
However, RPKI is vulnerable to forged origin attacks. In this type of attack, the adversary claims a non-existent link to the victim in a malicious BGP update. Since RPKI only validates the origin of the IP prefix in a BGP update, the malicious update will propagate even in the presence of ROV.

An AS can implement route filters on neighbors' BGP announcements to allow announcements only from approved IP prefixes or AS paths~\cite{MANRS, snijders2016peerlock}. However, strict prefix-based route filtering is difficult to scale to peer-to-peer links and larger networks with a substantial number of IP prefixes; AS-path filtering cannot prevent an adversary from announcing a malicious prefix with a legitimate-looking AS path.

\textbf{Secure Internet backbone candidates.}
A wide variety of secure Internet routing technologies can be used as a secure backbone,
ranging from BGP extensions to entirely new Internet architectures~\cite{kent2000secure,wan2005pretty,karlin2006pretty,xu2006miro,kushman2007rbgp,xu2007deft,motiwala2008path,liao2008reliable,wang2009path,camacho2013bgp,naylor2014xia, Perrig2017}.
Due to space limitations, only a few approaches are discussed below, but a more comprehensive overview of secure routing architectures is also available~\cite{butler2009survey,mitseva2018state,rexford2010future}.


\emph{Federated backbones.}
BGPsec offers the properties required for a secure backbone by
augmenting BGP to provide cryptographic verification of routes in
the control plane~\cite{austein2017rfc}. BGPsec requires each AS to
sign outbound BGP announcements, thus allowing ASes along
the path to verify the authenticity of the announcement. BGPsec not
only prevents ASes from falsely originating prefixes that were not
allocated to them, but also prevents ASes from claiming fake
adjacencies. Several shortcomings hampered wide-spread BGPsec deployment so far, e.g., scalability issues,
slow convergence, high overhead for update verification, and
vulnerabilities that remain unaddressed. 

New Internet architectures can also be used for constructing a federated backbone, such as NEBULA~\cite{anderson2013nebula}, NIRA~\cite{yang2007nira}, and  SCION~\cite{Zhang2011,Perrig2017}.
Specifically, SCION has been suggested as a clean-slate Internet architecture to provide secure inter-domain routing. SCION provides strong security properties: in-network per-packet source authentication, sovereignty and transparency for trust roots, and attack resilience for inter-domain routing.
Of these architectures, SCION is available today as a production network from several ISPs.

\emph{Private backbones.}
Several corporations have developed proprietary private backbones that
allow for secure data delivery, e.g., AWS and Cloudflare Argo~\cite{aws, argo}. While not
federated, some of these backbones allow for participants to connect
via VPN tunnels and even announce their own address space. While these
commercial offerings are promising, they are challenging to scale as
the competing providers do not seem to move towards a federated
offering. 
\vspace{-1mm}
\section{Design Principles}
\label{sec:overview}

\vspace{-1mm}
\subsection{Goals and Challenges}
\label{sec:goal}

We seek to design a secure routing architecture that provides to the Internet \textbf{high resilience against BGP hijacking attacks}. However, our intention is not to introduce another secure routing protocol. Although secure routing protocols provide clear advantages over the currently used BGP, they have so far only achieved partial adoption. The main obstacle to large-scale adoption is that a participating entity requires a sizable financial investment while gaining limited benefits at the early stage of deployment. 
To overcome this, the architecture has to be: (1) readily deployable without modifications to existing Internet infrastructure and protocols, (2) readily available for customers who want to use the system, 
requiring minimum changes in setup,
and more importantly, (3) readily beneficial to the customer even with a partial deployment of the architecture.
Considering incremental deployability, we aim to leverage an already-deployed secure routing infrastructure (as a secure backbone) to mediate communication between traditional IP endpoints. This extends the benefits of the secure backbone to the broader Internet, and kick-starts routing security for Internet communication. 
We call this system SBAS, the Secure Backbone AS. It is important to note that SBAS does not compete with any other secure routing methods; SBAS can benefit from them since it is a complementary system, improving security in a synergistic manner. From this approach, the following research challenges arise:

\textbf{Architectural continuity.}
Coupling of a secure routing infrastructure and the rest of the Internet requires architectural continuity. That is, the secure backbone understands BGP's control plane and seamlessly bridges remote BGP peers while leaving the leveraged secure routing infrastructure and its security guarantees intact. 
To this end, the secure backbone must achieve an architectural abstraction of the underlying infrastructure and provide a transparent interface to customers.

\textbf{End-to-end security.}
In the context of mediating customer's IP endpoints 
via a secure backbone, the end-to-end communication path can be segmented into: 
an external (insecure) segment, which is comprised of the Internet links between an IP endpoint and the SBAS ingress/egress point, and
an internal segment between an arbitrary ingress and egress pair of the secure routing infrastructure. 
To ensure end-to-end secure routing,  (1) a customer must be able to select trusted ingress/egress points and securely exchange packets with hijack resilience, and (2) the secure backbone must deliver the security properties it promised to any pair of ingress/egress points even in the presence of internal adversaries. 

\textbf{Routing priority.} 
To enable customers to route traffic from/to the Internet through a secure backbone, SBAS needs to disseminate the customers' prefix announcements to all other customers and external entities. Prefixes will then be announced via SBAS and the Internet, resulting in competing announcements. To maximize the ability to route securely, SBAS must be able to convince the entities receiving the announcements to prioritize routing paths through the secure backbone over the insecure Internet paths.




\vspace{-1mm}
\subsection{Threat Model}
\label{sec:threat-model}

\textbf{Adversary types.}
SBAS considers
two distinct types of adversaries.
The first type is an external adversary, who controls an AS on the Internet and is able to make arbitrary BGP announcements.
The adversary performs BGP attacks to hijack or intercept the traffic originated from or destined to customers, which enables more sophisticated attacks such as domain validation attacks~\cite{birge2018bamboozling} and traffic analysis~\cite{sun2015raptor}.
%
The second type is an internal adversary, who may compromise entities with various roles in SBAS, attempting to disrupt connectivity to legitimate customers.
%
We also allow the two types of adversaries to collude.
Attacks that do not target routing (e.g., exploiting implementation vulnerabilities, or DDoS attacks) are considered out of scope. 
 
\textbf{BGP attack types.}
The primary threat that SBAS aims to defend against is equally-specific prefix attacks. This is justified since all prefixes controlled by SBAS are announced as 24-bit prefixes (or 48-bit IPv6 prefixes), which can only be attacked via equally-specific prefix attacks (recall that prefix announcements longer than 24 bits in IPv4 and 48 bits in IPv6 are typically filtered). 
Even though the primary threat we considered is equally-specific prefix attacks, we show that
communication between two SBAS customers benefits from increased resilience even in the presence of more-specific prefix attacks. We demonstrate this property in \cref{sec:opt-in-peering}.
 

\vspace{-1mm}
\section{Design of SBAS}
\label{sec:design}

This section describes the control plane, data plane, and operational aspects in the design of SBAS.

\begin{figure}[t]
\centering
  \includegraphics[width=0.9\linewidth]{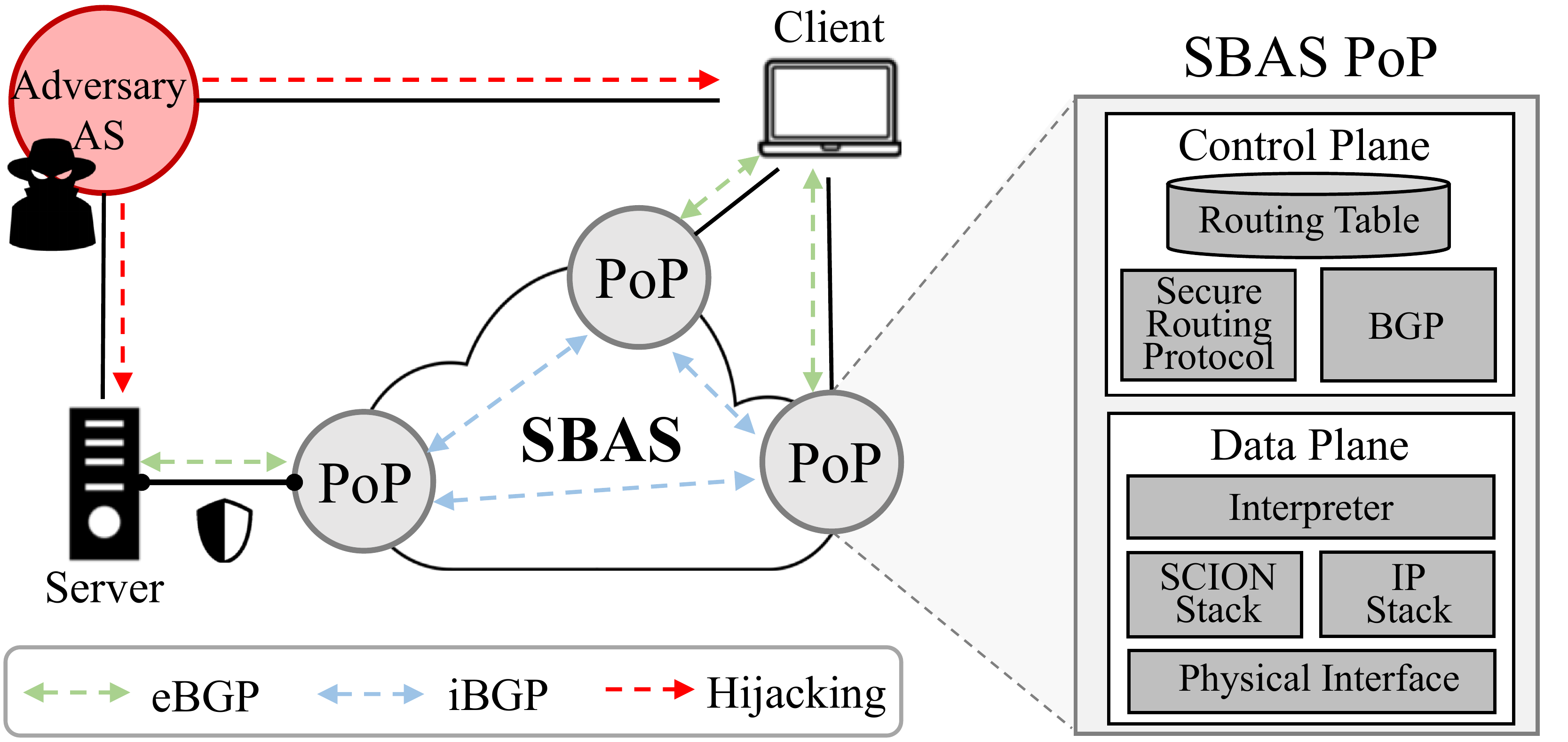}
  \caption{
    SBAS Overview
   %
  }
  \label{fig:overview}
\end{figure}

\vspace{-2mm}
\subsection{SBAS Overview}

SBAS is an abstraction that enables a federated backbone network to act as a single AS toward the outside Internet.
As shown in \cref{fig:overview},
customers of the system can connect via secure connections (e.g., VPN tunnels) to one or more PoPs, which are located at the edge of SBAS.
The system supports both
(1) customers that control their own address prefixes to be routed via SBAS, and
(2) customers that operate smaller network domains. The latter can simply obtain addresses from an SBAS-owned address range.
Internally, the PoPs form a full-mesh BGP topology over the internal routing protocol of the backbone, which is used to distribute customer announcements to the globally distributed PoPs and to achieve maximum security for traffic to secured prefixes.
SBAS is fully compatible with conventional BGP security practices and internally performs validation checks to ensure that only legitimate announcements are redistributed by the system.

Moreover, the secure routing protocol used internally, along with additional security mechanisms, ensure that the redistribution scheme tolerates misbehaving SBAS members.
This enables the system to extend the benefits of the secure federated backbone to the broader internet, while addressing the challenge of partial deployment incentives that are limiting the practical use of such approaches.


SBAS distinguishes between the following roles:

\textbf{Customer.}
A customer is an entity that resides outside the backbone and obtains service from SBAS through a contract, which enables it to route traffic securely through the system.
SBAS supports both (1) customers that only control single hosts
(e.g., server operators or end users), and (2) entities that own
entire address ranges and AS numbers.

\textbf{Point of Presence (PoP).}
A PoP is a member of SBAS that is located at the edge, i.e., provides connectivity to SBAS customers and interfaces with the regular Internet.

\textbf{Backbone operator.}
Such entities participate in the backbone network, but are not located at the edge; they simply participate in the internal routing and forwarding.
This type of member does not need to be aware of the SBAS
infrastructure running on top of the backbone.
    
\textbf{External entity.}
This term refers to entities on the Internet that are unaware of SBAS.

SBAS distinguishes among three address categories:

\textbf{Secure.} This includes prefixes announced by SBAS customers and
    SBAS-owned address ranges, which are assigned to
    customers. Secure address ranges are announced publicly via BGP.
    
 \textbf{Internal.} To provide an internal addressing scheme among PoPs, e.g., to set up iBGP sessions between PoP routers, the PoPs reserve
 address space for SBAS internal operation. This address
 space is not visible outside the SBAS infrastructure.
    
\textbf{Global.} We use this term to refer to all globally routable addresses to which the categories above do not apply.

\vspace{-2mm}
\subsection{The SBAS Abstraction}
\label{sec:design-abstraction}

Toward the Internet, SBAS is abstracted as a single AS making BGP announcements.
The defining characteristic of the system is that it employs a federated structure internally:
various entities may participate by connecting to the backbone network, which runs a secure inter-domain routing protocol.
Compared to offering secure routing through a tier-1 ISP or IXP that allows any customers to connect via a secure channel (e.g., VPN or direct physical link), SBAS's federated structure and abstraction provide the following benefits: 
(1) federation lowers the potential for centralization of the Internet and surveillance of traffic at the hyper-connected single node, 
(2) incremental deployability by allowing other ASes to participate, 
and (3) an expanding SBAS network results in a reduction of the hop distance to customers which increases resilience to routing attacks.

\paragraph{Virtualized full-mesh iBGP}
The internal structure of SBAS can be abstracted to a full-mesh
topology between the PoPs, independent of the routing
protocol of the backbone.
Over these connections, the PoPs redistribute announcements from SBAS customers as well as the Internet, akin to the operation of iBGP in a regular AS.
In order to prevent tampering by non-PoP members, the iBGP sessions run over an encrypted and authenticated connection (such as a VPN tunnel).

\vspace{-1mm}
\subsection{Secure Route Redistribution}
\label{sec:design-redistribution}

SBAS offers a high degree of flexibility to its customers through support for dynamic route redistribution.
Contrary to a traditional AS, which is controlled by a single entity, the redistribution scheme to be used in SBAS must support its federated structure and remain secure in the presence of malicious members.
In the following, we describe the design and security aspects of the route redistribution mechanism.

\begin{figure}
    \centering
    \includegraphics[width=0.85\linewidth]{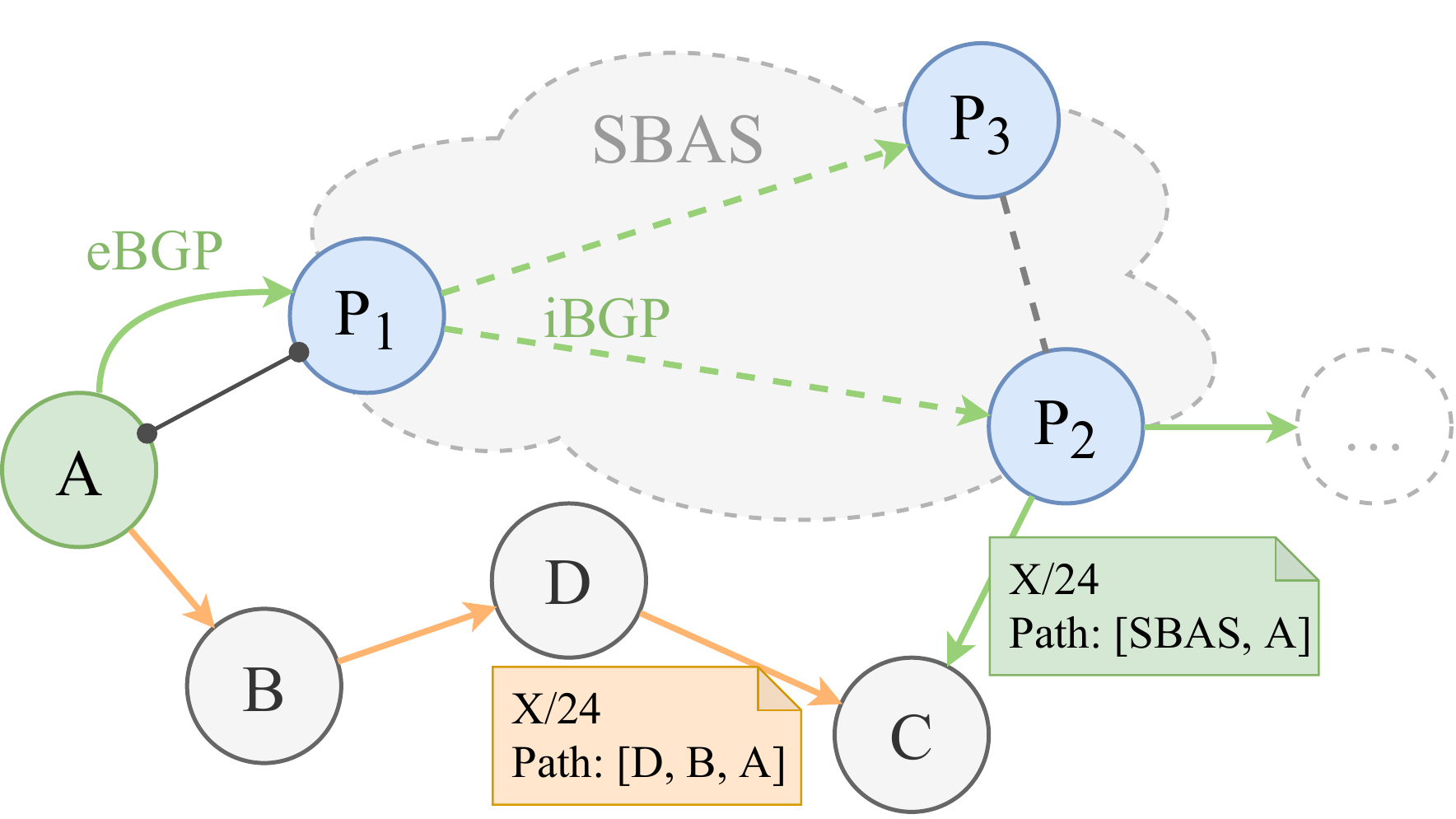}
    \caption{
        The route redistribution process for a prefix $X$ owned by customer AS $A$.
        The prefix is being announced in parallel through SBAS and to neighbors of $A$.
    }
    \label{fig:redistribution}
\end{figure}

\paragraph{Federated bring-your-own IP prefix}
Customers that already control one or multiple IP prefixes can use them directly with SBAS.
For this purpose, SBAS implements a route redistribution mechanism that enables a customer to route incoming traffic from the Internet through the secure backbone.
The process is depicted in \cref{fig:redistribution}:
the customer (AS $A$) initiates a BGP session with the PoP (AS $P_1$) over a VPN connection.
Using this session, $A$ makes an announcement for its prefix $X$,
which is then redistributed to all other PoPs over the full-mesh iBGP
topology.
A remote PoP such as $P_2$, upon receiving such an announcement over the iBGP
session with $P_1$, sends it out to its eBGP neighbors, i.e., Internet peers as well as SBAS customers.

\paragraph{Enhanced RPKI-based security}
The RPKI system provides strong security properties for the first hop of BGP advertisements, but does not protect subsequent hops.
The design of SBAS complements this property, as it eliminates attack surfaces on the path through its secure backbone.
SBAS leverages RPKI to defend against two distinct threats:
(1) customers advertising prefixes that they do not own, and
(2) PoPs falsely claiming authorization for a prefix from a customer.

The first threat 
is prevented in SBAS using route validation at the ingress.
Each announcement from a customer must carry a valid ROA that authorizes the AS to originate the prefix, which is verified both at the ingress PoP and by the other PoPs that receive the redistributed announcement.
To prevent sophisticated routing attacks, SBAS additionally verifies that the AS path of these announcements does not contain any ASNs other than that of the origin (but still allowing for customer traffic engineering using path pre-pending).

An example of the second type of threat would be $P_3$ (in \cref{fig:redistribution}) forwarding the announcement received from $P_1$ in an attempt to attract traffic to $A$.
To prevent such malicious behavior, a customer can use RPKI to authenticate a single or multiple PoPs that are authorized to re-distribute a given prefix.
This approach is similar to \emph{path-end validation}~\cite{cohen2016jumpstarting}, but in this case, it can be used purely by SBAS members and customers without requiring any external deployment.

\paragraph{SBAS-only prefix}
Using an SBAS-defined BGP community tag, the customer can instruct the PoPs to only redistribute the announcement internally, i.e., to connected customers.
This enables full protection of an address range against hijacking attacks, since secure prefixes are always prioritized by SBAS members and customers (as described in \cref{sec:design-routing}).

\vspace{-1mm}
\subsection{Customer Perspective}
\label{sec:customer}

\paragraph{Service management}
Customers sign up for SBAS via an interface (e.g., SBAS portal) by setting up a contract at their local PoP (details on governance aspects are given in \cref{ref:governance}).
The connection to SBAS is managed by a client software that receives information about existing PoPs, including publicly reachable IP addresses and a VPN public key for each PoP. 
SBAS can suggest a default SBAS PoP based on a network proximity metric.

\paragraph{SBAS connection setup}
A customer can connect to SBAS by connecting to one or multiple PoPs.
A customer looking to maximize resilience to BGP attacks should generally prioritize the PoP that is the fewest BGP hops away.
The connection to a PoP is set up over a VPN tunnel with the PoP's keypair, or where possible, the customer can connect directly at the PoP.
The latter case has the additional benefit of eliminating the possibility of any BGP attack on the connection, as a local layer-2 connection can be used.
If the customer uses multiple PoPs, one connection is designated as the primary ingress point, with the others serving as backups for improved failure resilience.
In order to prevent routing loops, the VPN endpoint that is used to connect to the SBAS PoP must be assigned a non-secure address. 
A customer may wish to designate a part of its address range to be routed via SBAS (advertised via SBAS), and separate this from their remaining address space (advertised normally on the Internet).

\paragraph{Secure address assignment}
For customers that do not control an address space,
SBAS can offer a (paid) feature to assign single addresses from a secure SBAS-owned prefix.
This option is configured via SBAS client software.
Upon assigning such an address to a customer, the PoP announces it to the other PoPs over the existing iBGP sessions.
This allows them to route traffic to the appropriate location and keep track of addresses that have already been assigned.

\vspace{-1mm}
\subsection{Routing Logic at PoPs}
\label{sec:design-routing}

The following sections describe how an SBAS PoP maintains routing information for the different types of addresses, and how packets are forwarded accordingly.

\vspace{-1mm}
\def\nexthop{\textsf{nexthop}}
\bheading{Control plane: Virtualized iBGP.}
As shown in \cref{fig:control-plane}, each PoP maintains several iBGP sessions to other PoPs over the backbone network, as well as eBGP sessions to customers and Internet peers.
From the information received over these sessions, two routing tables are constructed:
The first table, which is given the highest priority, maps secure addresses to internal addresses.
Each entry may be either a remote customer prefix that is mapped to an internal address representing another PoP (in the example, \textsf{1.0.0.0/24} to $P_2$), or a local customer prefix that can be delivered to the customer's VPN endpoint (\textsf{2.0.0.1/32} to $\nexthop(D)$).
The advertisements for such routes are received over the iBGP session from other PoPs (in the former case) or over the eBGP session from customers (in the latter case).

As a lower-priority table, $P_1$ maintains an Internet routing table for routes obtained from its Internet peers.
These routes will likely also be received via iBGP from other PoPs distributing prefixes they received from their respective neighbors.
In this case, route selection can follow standard Internet policies or custom logic implemented by $P_1$.

\begin{figure}
    \centering
    \includegraphics[width=0.85\linewidth]{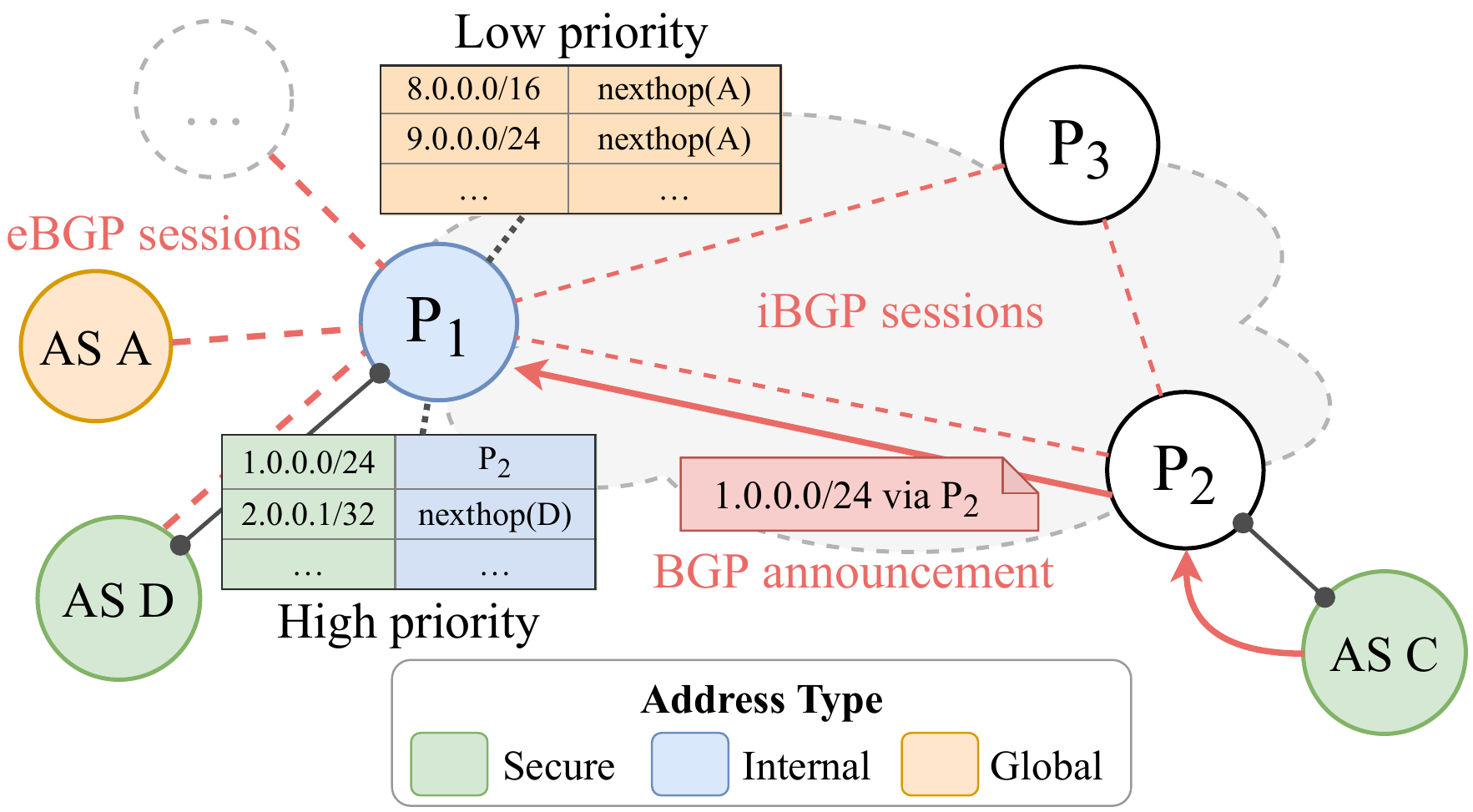}
    \caption{
        Control-plane configuration of $P_1$.
        $\nexthop(X)$ refers to the address used as the next hop to AS $X$.
    }
    \label{fig:control-plane}
\end{figure}

\vspace{-1mm}
\bheading{Data plane: Secure address prioritization.} 
\label{sec:priority}
Next, we describe data-plane forwarding decisions for different scenarios of source and destination locations, as illustrated in \cref{fig:routing-ingress}.

By keeping a strict priority hierarchy between secure routes and external routes, SBAS provides resilience to BGP hijacking attacks by design.
The detailed security properties achieved by this design are explored in \cref{sec:hijacking-experiments}.

\begin{figure}
    \centering
    \includegraphics[width=0.85\linewidth]{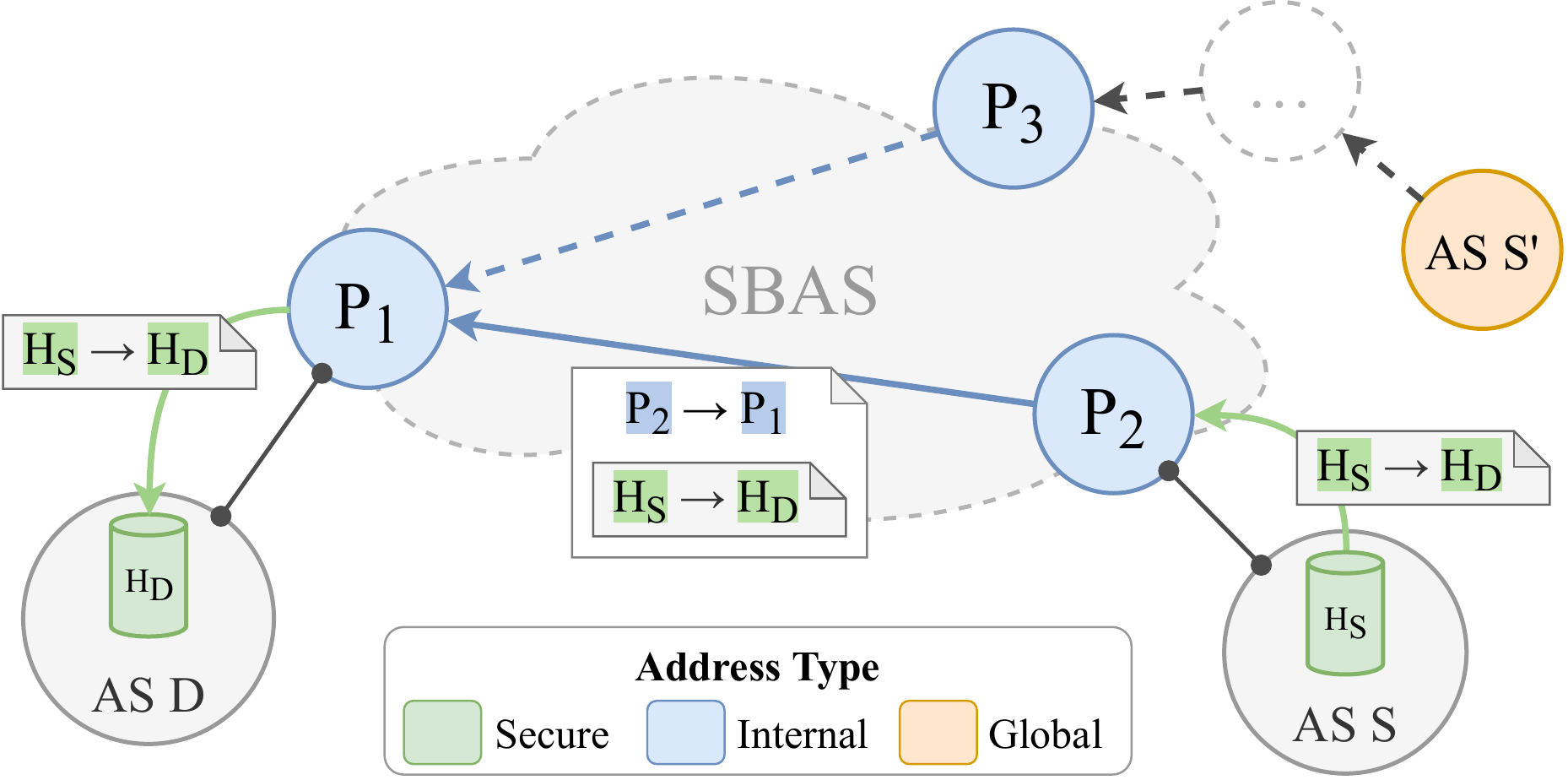}
    \caption{
        Routing logic for incoming traffic to a customer $D$ who owns a secure address range.
        The packet shown is sent from a secure address $H_S$ and encapsulated across SBAS using the internal addresses for the PoPs $P_1$ and $P_2$, before it is delivered to the secure address $H_D$.
        Packets may also originate from global addresses such as from AS $S'$.
    }
    \label{fig:routing-ingress}
\end{figure}

\emph{Customer-to-customer} ($H_S \rightarrow H_D$)
In the simplest case, a packet originates from a secure address $H_S$ in a customer AS $S$ and is destined to another secure address $H_D$.
The packet from $H_S$ is routed through the VPN tunnel to the ingress PoP $P_2$.
There, $P_2$ looks up the secure address $H_D$ and finds the internal address for the egress PoP $P_1$ associated with it.
The original packet is encapsulated over the backbone's internal protocol to $P_2$, which delivers it across the VPN tunnel to the destination in AS $D$.

\emph{External origin} ($S' \rightarrow H_D$)
We consider a packet that is destined to the secure address $H_D$, but originates from a source in AS $S'$ that is unaware of SBAS.
In this case, the data plane operations follow the same sequence:
Having received a BGP announcement from $P_3$ for the secure prefix that contains $H_D$, AS $S'$ will forward the packet to the nearest SBAS PoP.
From this point, the same logic is applied as in the previous case.

\emph{External destination} ($H_D \rightarrow S'$)
For traffic with global destination addresses, the routing decision offers more options through the choice of the egress PoP.
Whereas in the previous cases, the traffic was directed to the destination's preferred PoP, the decision to select an egress location is up to the ingress PoP.
This makes it possible to optimize for different metrics such as hijack resilience, which can be achieved by routing based on shortest AS path length, combined with hijack detection.
For instance, if an egress point notices a recent change in the AS origin of an IP prefix, this egress point may be avoided for traffic to this IP prefix until the origin change can be validated.
This approach further improves the system's resilience to external BGP hijacking attacks.

Through a number of key design principles and by leveraging the secure backbone for internal routing, SBAS is able to disseminate routes securely to customers and out to the Internet.
Using a strict priority hierarchy on the control plane, traffic to/from customers benefits from strong hijack resilience.

\vspace{-2mm}
\section{Implementation and Deployment}
\label{sec:implementation}
\vspace{-1mm}

We have implemented the SBAS design and deployed it on a globally distributed infrastructure. 
A key feature of our implementation is that it minimizes the need for new software and composes existing networking components in a synergistic manner. 
Driven by this, we implement the prototype of the SBAS system on top of the globally distributed future Internet research network SCIONLab.
The SBAS system is comprised of four PoPs, two PEERING announcement nodes, and three customer locations (\cref{app:node-location} presents the deployment map).
Our SBAS implementation running on the PoP consists of approximately $\num{1000}$ lines of code.
The software automatically configures and runs the various PoP components based on configuration files that describe the setup of the SBAS instance.
The full source code is publicly available.

\textbf{Instantiating SBAS with SCION.}
%
We employ SCION as the secure internal routing architecture of SBAS for several reasons:
(1) SCION already provides a strong PKI system by design, which is essential for the core SBAS properties such as secure route redistribution, heterogeneous trust for the federated participants, and cryptographic protection in routing,
(2) SCION provides network programmability with a high degree of freedom, helping us to virtualize the internal network structure and build a full-mesh intra-SBAS topology, and
(3) SCION possesses sufficient system maturity with real-world deployment and operation. 

To integrate SBAS with the existing SCION architecture, we leverage the SCION-IP Gateway (SIG), which translates between IP and SCION through en- and de-capsulation of packets.
The operation of a SIG at each PoP provides transparent IP connectivity without requiring any changes to customers' networking stacks.
%
We construct the SBAS protoype deployment on SCIONLab~\cite{kwon2020scionlab}, the global SCION research network 
spanning over 50 infrastructure ASes across the world.
In our deployment, four ASes instantiated at AWS datacenters in Oregon, Frankfurt, Singapore, and Tokyo are directly connected to the SCIONLab core infrastructure, operating as SBAS PoPs.



\begin{figure}[tbp]
    \centering
    \includegraphics[width=0.9\linewidth]{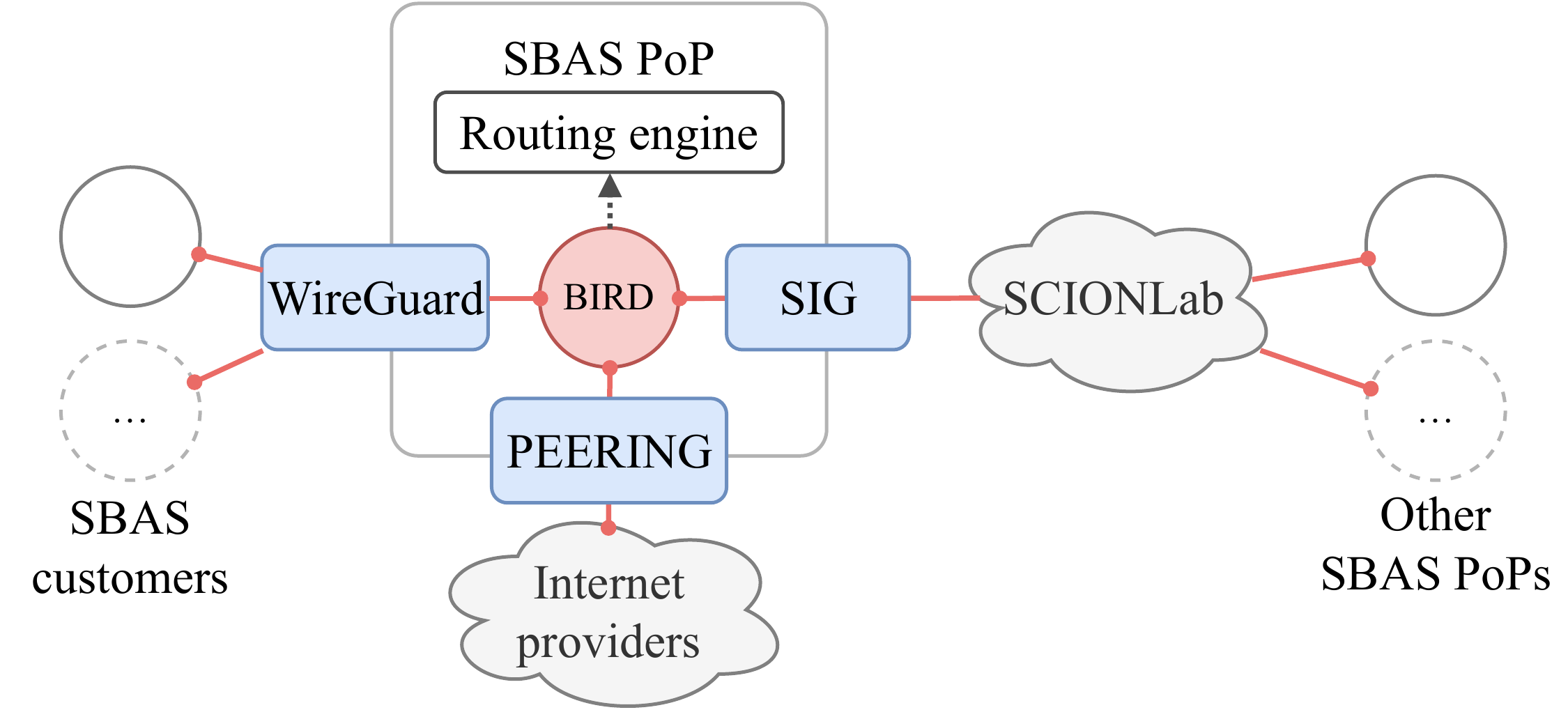}
    \caption{The interfaces (shown in filled rectangles) and BGP sessions (solid lines) maintained by an SBAS PoP.}
    \label{fig:pop}
\end{figure}

\textbf{Data-plane interfaces.} Each PoP has three interfaces for different types of destinations, as shown in \cref{fig:pop}: (1) A WireGuard instance to send/receive packets to/from SBAS customers that are connected to that PoP, (2) A traditional Internet interface with IP transit/peering and a BGP routing table, and (3) A SIG that encapsulates IP packets in SCION packets and sends them over the SCION backbone. This modular decomposition of interfaces enables a high degree of flexibility for SBAS.
For instance, a different backbone architecture can be configured to replace SCION as a drop-in replacement without requiring changes to the other parts of the PoP software.

\textbf{Control-plane management.} In addition to these data plane interfaces, each SBAS PoP maintains BGP sessions with customers, IP transit providers/peers, and other SBAS PoPs. These BGP sessions are handled by the BIRD Internet routing daemon~\cite{bird}. However, BIRD does not make the final routing decision; it simply exports the routes learned from its various BGP sessions into routing tables, which are then processed with different priorities by the SBAS routing engine. More details are presented in~\cref{sec:appendix:details}.

\textbf{Routing engine.} The SBAS routing engine compiles the routes from these BGP sessions and produces the final forwarding table that enforces the security/route preference requirements of SBAS. In addition to enforcing that secure SBAS customer routes are used over standard Internet routes, the SBAS routing engine can be extended to consider which SCION paths are used to reach specific SBAS PoPs, enabling advanced route selection models such as carbon-emission-based routing~\cite{carbon_routing}. See~\cref{sec:appendix:details} for more details.

\textbf{Packet handling.}
Communications inside SCION (e.g., from the SIG interface to remote PoPs) do not use IP for addressing. In SBAS, we encapsulate a SCION packet into an IP packet, using Generic Routing Encapsulation~\cite{RFC1701}, and maintain a \emph{static} table at the SIG that contains a single entry per PoP, mapping its internal IP address to its SCION address. 
This  enables us to have a unified IP routing table for both IP and non-IP packets. The actual routing of each packet can be performed efficiently by the Linux kernel.

\textbf{BGP connectivity.}
To provide PoPs with BGP connectivity and an Internet routing table, we use the PEERING framework~\cite{schlinker14peering}, which allows researchers to make BGP announcements and forward packets through peers and upstream transit providers.
Using this component, our announcements are propagated to the worldwide BGP ecosystem.

\vspace{-2mm}
\section{Latency Evaluation}
\label{sec:eval}
\vspace{-1mm}

Using the global deployment illustrated in \cref{fig:eval-topology}, we conduct a series of experiments measuring the latency achieved by SBAS in realistic scenarios.
SBAS is competitive with Internet latency and even improves upon it in some cases, despite running in a research testbed using mostly overlay links. Since the PoP components only introduce sub-millisecond overhead~(\cref{app:latency}),  
we focus on SCIONLab overhead and end-to-end latency in this section. 

The final end-to-end latency $\ell_{S \rightarrow D}$ between a source customer $S$ (connected to ingress PoP $I$) and a destination customer $D$ (connected to egress PoP $E$) is composed as follows:
\begin{equation*}
    \underbrace{
        \ell_{S \rightarrow D} =
        \ell_{S \rightarrow I} +
        \ell_{E \rightarrow D} +
        \overbrace{
            2 \cdot \textit{delay}_{\textit{PoP}}
        }^{\text{\cref{app:latency}}} +
        \overbrace{
            \ell_{I \rightarrow E}
        }^{\text{\cref{sec:latency-nodes}}}
    }_{\text{\cref{sec:latency-optin}}}
\end{equation*}

\subsection{Latency Optimization Between PoPs}
\label{sec:latency-nodes}

We demonstrate that the backbone network can be leveraged to optimize and even reduce latency between PoPs. 
By offering a choice between multiple paths, SCION enables applications to choose optimal paths based on various metrics, whereas BGP by design always selects a single path between each $\langle$source, destination$\rangle$ pair (which in many cases could result in sub-optimal latency~\cite{kwon2020scionlab}).

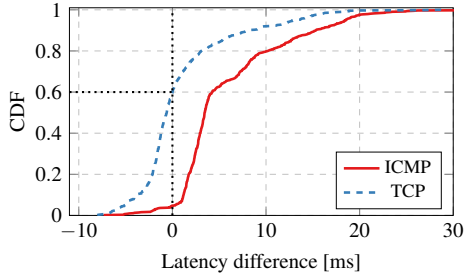
\begin{figure}[t]
\centering
\resizebox{0.75\linewidth}{!}{
\begin{tikzpicture}
\begin{axis}
[
every axis plot/.style={
line width=1.25pt,
},
cycle list/Set1,
cycle multiindex* list={
Set1
    \nextlist
[3 of]linestyles
    \nextlist
},
width = 0.95\linewidth, 
height = 0.6\linewidth,
xlabel = {Latency difference [$\si{ms}$]},
ylabel = CDF,
xlabel style={name=xlabel},
ylabel style={name=ylabel, yshift=-1em},
ymin = 0,
xmin = -11,
xmax = 30,
ymax = 1.01,
grid = major,
grid style={dashed, gray, opacity=.5},
legend pos = south east,
legend style = {font = \small},
]

\addplot table[ x index = {0}, y index = {1}, col sep=comma] {figures/data/latency-nodes-icmp.csv};
\addplot table[ x index = {0}, y index = {1}, col sep=comma] {figures/data/latency-nodes-tcp.csv};

\legend{ICMP, TCP}
\addplot +[mark=none, color=black, line width=1, dotted] coordinates {(0, 0) (0, 1)};
\addplot +[mark=none, color=black, line width=1, dotted] coordinates {(-11, 0.6) (0, 0.6)};
\end{axis}
\end{tikzpicture}
}
\caption{ Difference between SCIONLab and Internet latency between all pairs of PoPs.
        A negative difference indicates that SCIONLab achieves better latency than the Internet.
        The individual latency measurements (before computing the difference) have an overall mean of $\SI{149}{ms}$.}
\label{fig:latency-nodes}
\end{figure}
\vspace{-1mm}
We measure latency between all pairs of PoPs across both SCIONLab (used in our SBAS deployment) and the Internet.
These measurements are compared in \cref{fig:latency-nodes}.
For the Internet latency baseline, we used different measurement methods to 
simulate real-world traffic, eliminating protocol-specific factors:
echo requests from the ICMP protocol, and TCP handshakes.
The SBAS latency is measured over SCION. 
Note that, although the same packet generator is used for the Internet latency and SBAS latency measurements, in SBAS it appears to the data center network as generic UDP traffic to locally deliver SCION packets.~\kwon{please check if we actually use the same traffic.}
We observed that the latency is consistently lower when measured using ICMP as opposed to the TCP-based measurements.
Following the methodology of Kwon et al.~\cite{kwon2020scionlab}, we use TCP as the point of comparison for SCION latency.

The results in Figure~\ref{fig:latency-nodes} show that \textbf{SCIONLab achieves lower latency across PoP nodes than the Internet for approximately $\mathbf{60\%}$ of the measurements}, despite consisting largely of overlay links.
The improvement stems from the sophisticated path control that SCION provides; SCION can steer packets through latency-optimized paths (e.g., Tokyo-Singapore-Frankfurt) while BGP selects a detour path (e.g., Tokyo-Seattle-Frankfurt).
This indicates the potential for improved latency using an inter-domain routing architecture like SCION, which is able to leverage its advantage even in this setting with relatively few path choices, and can compensate for the overhead of tunneling 
to create overlay links.

\vspace{-1mm}
\subsection{End-to-End Latency}
\label{sec:latency-optin}

The final latency experiment evaluates the performance achieved for end-to-end connections between two customer hosts that communicate across SBAS. 
To evaluate this performance, we ran one round of the latency measurement 
between all pairs of customer machines in our testbed every hour over the span of two weeks (336 rounds in total).  
In each round, we ran ping 30 times with a one-second interval and computed the averaged  latency of ping packets. 
We picked three customer locations across multiple continents and chose the closest PoP as the ingress for each customer.
Therefore, these measurements are expected to compare less favorably to the Internet baseline than the latency over SBAS, as traffic may need to take a detour from the source over the nearest ingress PoP and from the egress PoP to the destination.
The results shown in \cref{fig:latency-optin} confirm this hypothesis and align with the measured latency from customers in the testbed to the nearest PoP, which is displayed in \cref{fig:latency-pops}.
\textbf{On average, end-to-end latency over SBAS is approximately $\mathbf{17}\,\mathbf{ms}$ higher than over the Internet.}
The end-to-end differences between SBAS and the Internet are relatively minor when the large intercontinental latencies are taken into account:
for instance, latency between Zurich and Osaka has a mean of $\SI{241}{ms}$ over the Internet and $\SI{259}{ms}$ over SBAS, which amounts to a relative increase of less than $7.5\%$.

\begin{figure}[t]
\centering
\resizebox{0.75\linewidth}{!}{
\begin{tikzpicture}
\begin{axis}
[
every axis plot/.style={
line width=1.25pt,
},
cycle list/Set1,
cycle multiindex* list={
Set1
    \nextlist
[3 of]linestyles
    \nextlist
},
width = 0.95\linewidth, 
height = 0.6\linewidth,
xlabel = {Latency difference [$\si{ms}$]},
ylabel = CDF,
xlabel style={name=xlabel},
ylabel style={name=ylabel, yshift=-1em},
ymin = 0,
xmax = 30,
ymax = 1.01,
grid = major,
grid style={dashed, gray, opacity=.5},
legend pos = north west,
legend style = {font = \footnotesize},
]

\addplot table[ x index = {0}, y index = {1}, col sep=comma] {figures/data/latency-optin-california-zurich.csv};
\addplot table[ x index = {0}, y index = {1}, col sep=comma] {figures/data/latency-optin-osaka-zurich.csv};
\addplot table[ x index = {0}, y index = {1}, col sep=comma] {figures/data/latency-optin-california-osaka.csv};
\legend{N. California -- Zurich, Osaka -- Zurich, N. California -- Osaka}
\addplot +[mark=none, color=gray, line width=1, dotted] coordinates {(0, 0) (0, 1)};
\end{axis}
\end{tikzpicture}
}
\caption{Difference of SBAS versus Internet latency between customers.
        The individual latency measurements (before computing the difference) have an overall mean of $\SI{154}{ms}$.}
\label{fig:latency-optin}
\end{figure}
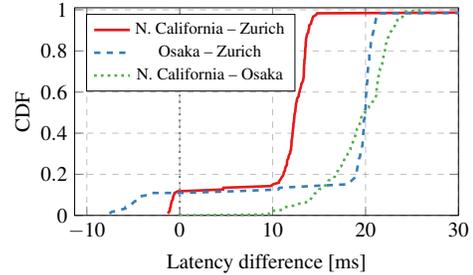
\vspace{-1mm}

In some scenarios, the latency improvements offered by SCIONLab even enable end-to-end connections to achieve better latency than the Internet.
The large variance in this difference over time can be attributed to the instability of BGP routes, which change frequently over time.

By using an expanded network of PoPs (reducing the latency from \cref{fig:latency-pops}) as well as dedicated SCION links in the backbone network (as mentioned in the previous section), SBAS's end-to-end latency can be improved further.

\vspace{-1mm}
\subsection{Discussion: Scalability}
\label{sec:scalability}
\vspace{-1mm}

Several aspects of SBAS's design enable it to scale to a large, real-world deployment. 
The bandwidth and compute-memory expense of SBAS is roughly proportional to the number of its clients. The PoP module (which includes the WireGuard tunnel, SIG, and routing logic) maintains no per-flow state. Adding a client to SBAS involves only adding a few internal routing table entries and the client's IP/VPN key to the PoP WireGuard configuration. SBAS capacity can be scaled to serve additional customers by increasing the computational capacity of the PoPs and increasing the bandwidth of the secure internal inter-PoP network. Key management and exchange between PoPs is handled by SCION, which improves upon RPKI scalability by delegating trust roots and key management to a limited number of isolation domains \cite{Perrig2017}.

We recognize that the growing size of Internet routing tables is a present concern \cite{routing_table_scaling}. Although SBAS infrastructure announces /24 prefixes (or /48 for IPv6 infrastructure, the longest publicly routable prefix in IPv6 \cite{ipv6_prfx_length}) for sub-prefix hijack protection, the number of such announcements is overall few and proportional to the number of SBAS POPs combined with the number of connected customer ASes (as each PoP or customer needs one prefix for their WireGuard endpoint). While SBAS customers are encouraged to use /24 length-prefixes (for select security-critical services), SBAS does not disaggregate customer announcements. 

\begin{table}[tb]
    \centering
    \footnotesize
    
    \begin{filecontents*}{figures/plots/latency-pops.csv}
Customer,PoP,Latency
Zurich,Frankfurt,12.364
N. California,Oregon,25.38
Osaka,Tokyo,12.745
\end{filecontents*}
\pgfplotstabletypeset[
        col sep=comma, 
        display columns/0/.style={
            column name={\textbf{Customer}},
            column type={l}, string type
        },
        display columns/1/.style={
            column name={\textbf{PoP}},
            column type={l}, string type
        },
        display columns/2/.style={
            column name={\textbf{Latency [$\si{ms}$]}},
            column type={r}, numeric type,
            precision=2, set thousands separator={}, fixed, fixed zerofill
        },
        every head row/.style={
            before row={\toprule},
            after row={\midrule},
        },
        every last row/.style={after row=\bottomrule},
    ]{figures/plots/latency-pops.csv}
    \caption{
        Latency from customers 
        to the respective closest PoP.
    }
    \label{fig:latency-pops}
\end{table}

\vspace{-2mm}
\section{Security Analysis}
\label{sec:analysis}
\vspace{-1mm}

We used two primary methods to evaluate the security of SBAS: (1) real-world attacks using the PEERING BGP research framework~\cite{schlinker2019peering}, and (2) simulated attacks using Internet topology simulations~\cite{gao2001gao_rexford}. These two methods are intended to complement each other. The PEERING framework allows us to launch ethical BGP attacks against the real SBAS prototype deployment and accurately captures the dynamics of Internet routing, but does not let us experiment with many different adversary locations (as we are limited by the nodes of the PEERING framework). 
Topology simulations allow us to experiment with a large number of adversary locations but yield less accurate results~\cite{juen2014defending}. However, when considered side-by-side, the results of these two types of evaluations complement each other, allowing for a more accurate understanding of the security of SBAS.

Recall from the threat model (Section \ref{sec:threat-model}) that we focus our evaluation on equally-specific prefix attacks because more-specific prefix attacks can be avoided by providing services through disaggregated IP prefixes, which we use for all SBAS-related addresses. We do additionally consider more-specific prefix attacks 
in Section \ref{sec:opt-in-peering}.

\vspace{-1mm}
\subsection{Evaluating SBAS Deployment Against Ethical Real-World BGP Attacks}
\label{sec:hijacking-experiments}

We launched real-world BGP attacks against our SBAS deployment using the PEERING framework. First, we set up two PEERING locations (known as \emph{muxes}) to act as SBAS customers.
One customer location was the mux utah01 located at the University of Utah. The other one was the mux grnet01 located in the Greek education and research network GRNET. Next, we used the mux neu01 at North Eastern University to serve as an adversary and launch BGP attacks.

\textbf{Ethical considerations.}
An important ethical principle underlying our experiment setup is that we only made BGP announcements for prefixes that we are authorized to use. 
%
Even though we used the neu01 mux as an adversary to model our attacks, it had proper authorization from the PEERING framework for all of its BGP announcements. 
Further, we used IP prefixes explicitly delegated to our infrastructure (from PEERING and participating educational institutions) that hosted no production services and served no real users. Finally, we also followed the PEERING framework's acceptable use policy as to not overwhelm or crash internet routers.

Our experiments validate that SBAS can mitigate both non-adaptive attacks that target customer prefixes as well as adaptive attacks that target customer tunnels to the SBAS PoPs. Moreover, SBAS can even enhance communication security between a SBAS customer and external hosts on the Internet. 
\cref{fig:peering-results} depicts a summary of our results.

\textbf{Control case: Successful attack 
without SBAS.}
We started with a control case where SBAS was not used and the customers in utah01 and grnet01 announced their prefixes via their traditional BGP providers/peers. 
The adversary node at neu01 then attempted to hijack communication between the two customers by announcing utah01's prefix and grnet01's prefix.  
We found that while traffic from grnet01 to utah01 was routed successfully to utah01, \textbf{traffic from utah01 to grnet01 was routed to the adversary}. This let the adversary observe (and potentially modify)  packets sent from utah01 to grnet01. Thus, the connection between utah01 and grnet01  was successfully attacked by the adversary in the absence of SBAS because utah01 had a superior BGP route to the adversary to the one it had to the
\ifnum\techreport=1
victim (see Section \cref{app:route-decisions}).
\else
victim.
\fi

\begin{table}[t]
    \centering
    \footnotesize
    \begin{tabular}{rcc} 
\toprule
\textbf{Attack target} & \textbf{No SBAS} & \textbf{Using SBAS} \\
\midrule
utah01 prefix & Failed & Failed \\
grnet01 prefix & \textbf{Succeeded} & Failed \\
utah01 tunnel & N/A & Failed \\
grnet01 tunnel & N/A & Failed \\
\bottomrule
\end{tabular}
\caption{
        Summary of ethically conducted real-world attack experiments (via PEERING) from an adversary at neu01.
    }
    \label{fig:peering-results}
\end{table}

\textbf{Attack mitigation using SBAS.}
\label{sec:opt-in-peering}
Next, we connected the utah01 and grnet01 customers to the prototype SBAS implementation using the Oregon and Frankfurt SBAS PoPs respectively. We then had utah01 and grnet01 make the announcements for their IP prefixes through SBAS. Recall that at each customer node, SBAS-learned routes are given higher priority (\cref{sec:priority}) than standard Internet routes.
We consider two types of adversaries: 

\textit{Non-adaptive adversary.}
The non-adaptive adversary is not aware of SBAS and launches BGP attacks against customers' IP prefixes as usual, as in the control case.
Because both utah01 and grnet01 were communicating through SBAS,
\textbf{the adversary was incapable of hijacking any of the traffic between grnet01 and utah01} in either direction of communication. Note that this result is independent of customer or adversary location. Two customers will always successfully resist BGP attacks where an adversary targets a customer's IP prefix announced through SBAS. This is due to route prioritization and holds even in the case of more-specific BGP attacks. SBAS PoPs load secure routes into a separate routing table that is given higher priority than the Internet routing table. 
SBAS customers' outbound traffic will always go through the connected SBAS PoPs. More-specific routes in the global routing table do not affect routes between SBAS customers.

\textit{Adaptive adversary.}
%
%
An adaptive adversary who is aware of SBAS may instead chose to attack the tunnels that each customer uses to communicate with SBAS.%
While this adaptive attack is inherently less devastating because tunnels are end-to-end encrypted,
there are still powerful attacks that can be launched against encrypted traffic~\cite{sun2015raptor, panchenko2011website}.

Using our PEERING setup, we had the adversary at neu01 attack the IP prefixes used by utah01 and grnet01 to establish their WireGuard sessions with SBAS.%
\footnote{Another attack can be launched by hijacking the IP prefix of the SBAS PoPs for their WireGuard endpoint. We could not conduct this  in the wild, as we were not authorized to hijack the SBAS PoPs' AWS-controlled prefix.} 
Even when the adversary maliciously announced the IP prefix of the WireGuard endpoint of both victims with an equally-specific BGP attack,
\textbf{communication between utah01 and grnet01 was uninterrupted and was never routed to the adversary.} Note that SBAS infrastructure prefixes, like the one used for the VPN endpoint, are required to be $/24$s; more-specific attacks against SBAS VPN endpoints are not viable.

We note that the success of this type of adaptive adversary against SBAS depends highly on the customer's choice of ingress points. 
As a contrived example, had utah01 chosen the Frankfurt SBAS PoP as its ingress point and grnet01  chosen Oregon as its ingress point, communication along both of the tunnels would have been routed to the adversary. It is because of the proximity of the SBAS ingress point to the SBAS customer (relative to the adversary's location) that SBAS offers improved security even against this type of adaptive adversary. 
In the optimal case, a customer may even be able to obtain a direct layer-2 connection with a PoP, thwarting this attack entirely.

\textbf{Characterizing communication security with external hosts.}
\label{subsec:experiment-method}
In addition to running experiments to measure the security of communication between SBAS customers in utah01 and grnet01, we evaluated the security benefit that SBAS offers when a SBAS customer is communicating with an non-SBAS-protected (i.e., external) host or server on the Internet. 
We build upon the methodology presented by Birge-Lee et~al.~\cite{birgelee2019sico}. We constructed a sample of 1k IP addresses from the Censys Internet-wide IPv4 scans~\cite{durumeric2015censys} to serve as external hosts. The sample was chosen at random 
and filtered to only include hosts that responded to ICMP echo (ping) requests.

To measure the impact of SBAS on communication between a SBAS customer and external hosts, we performed ethical equally-specific prefix hijacks from the adversary neu01 against IP prefixes originated by utah01 and grnet01 and announced by SBAS (using nodes at amsterdam01 and seattle01). Similarly, we also performed ethical hijacks where no SBAS was used as the control case. 
Then, for each hijack, we launched a ping scan of the 1k external hosts in our sample from an IP address in the prefix that was being hijacked. When each host in the sample received the ping request, it generated a ping response with a destination IP address that was under attack by the adversary. Computing the fraction of hosts whose ping responses were routed to the adversary allowed us to measure the impact of each hijack we launched.

\textit{SBAS significantly enhances communication security with external hosts.} When the utah01 and grnet01 nodes were not using SBAS, the adversary at neu01 was able to hijack traffic from 72\% and 76\% of the 1k external hosts respectively. When we connected utah01 and grnet01 to SBAS and launched an attack on the SBAS announced-prefix, \textbf{the adversary's hijacking capability reduced threefold to affecting only 25\% of hosts.} We emphasize that the presented security improvements are conservative, as this experiment was performed against the SBAS prototype deployment that uses only the seattle01 and amsterdam01 PEERING muxes. We are actively working to expand this deployment, and present recommended expansion steps and the associated security improvement in \cref{sec:improved-resilience}. As more nodes are added, hosts will 
have a shorter route to the nearest SBAS PoP which will further reduce the spread of the adversary's attacks.

\subsection{Quantifying Hijack Resilience via BGP Attack Simulations}
\label{sec:simulation}

To evaluate SBAS security beyond PEERING mux adversaries and client locations, we employ Internet-scale attack simulations. The Internet topology was constructed using the CAIDA AS topology dataset \cite{caida}, augmented with peering information inferred from the bdrmap tool~\cite{bdrmap} and BGP Routing Information Base (RIB) data collected from Route Views~\cite{routeviewUO} and RIPE NCC RIS~\cite{ripe_ncc_ris} to correctly model route selection at the AWS datacenters and PEERING nodes that are part of the SBAS deployment. We build upon the methodology developed by Birge-Lee et~al.~\cite{birge2021le} to perform prefix-level (as opposed to AS-level) simulations.
We evaluate how likely traffic from external sources will still be routed to SBAS (via the PEERING framework) in the event of an equally-specific BGP attack.
We also explore how security improves with
more BGP-announcing SBAS nodes and with full deployment of RPKI in the broader 
internet. 


\textbf{Computing attack resilience metric.}
We use the notion of \emph{resilience}~\cite{wubbeling2016res} to quantify the fraction of total potential adversaries that are topologically unable to launch 
an equally-specific prefix hijacking attack for an arbitrary prefix announced by a victim AS. The detailed definition of resilience is in \cref{app:sim:resdef}.
%
%
Resilience is affected by the relative locations of the victim and adversary ASes, their peering and provider relationships, and the application of further security measures, including RPKI.
\ifnum\techreport=1
(See \cref{app:route-decisions} for a more detailed discussion of the factors which influence the selection and hijacking of routes.)
\fi

\paragraph{Internet topology simulations}
We run BGP simulations against an AS-level adversary considering a random sample of 1k adversary ASes as the attacker set $\mathcal{A}$, corresponding to approximately $1.39\%$ of the $N = 71669$ ASes profiled in the CAIDA AS topology. Against each attacker AS, we consider all other $N - 1$ ASes as the set of external hosts (traffic sources) $\mathcal{B}$.
Given these fixed sets of adversary ASes and external hosts, we run BGP simulations for two scenarios: (1) a victim prefix is announced via SBAS BGP announcement nodes; (2) a victim prefix is announced in a conventional manner without SBAS. We consider a random sample of 1k victim prefixes, which are selected based on the methodology in \cref{subsec:experiment-method}.
We evaluate SBAS configurations with varying number and location of BGP announcement nodes. When varying the number of BGP-speaking SBAS nodes, we present results for configurations (node locations) that are globally optimal (more details in \cref{app:node-location}). 

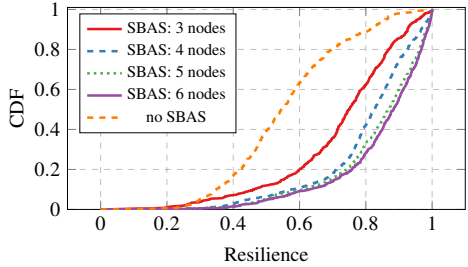
\begin{figure}[t]
    \centering
    \resizebox{0.75\linewidth}{!}{
    \begin{tikzpicture}
\begin{axis}
[
every axis plot/.style={
line width=1.25pt,
},
cycle list/Set1,
cycle multiindex* list={
Set1
    \nextlist
[3 of]linestyles
    \nextlist
},
width = \linewidth, 
height = 0.6\linewidth,
xlabel = {Resilience},
ylabel = CDF,
xlabel style={name=xlabel},
ylabel style={name=ylabel, yshift=-1em},
ymin = 0,
ymax = 1.01,
grid = major,
grid style={dashed, gray, opacity=.5},
legend pos = north west,
legend style = {font = \footnotesize},
]

\addplot table[ x index = {0}, y index = {1}, col sep=comma] {figures/data/simul_cdf_sbas_default.csv};
\addplot table[ x index = {2}, y index = {3}, col sep=comma] {figures/data/simul_cdf_sbas_default.csv};
\addplot table[ x index = {4}, y index = {5}, col sep=comma] {figures/data/simul_cdf_sbas_default.csv};
\addplot table[ x index = {6}, y index = {7}, col sep=comma] {figures/data/simul_cdf_sbas_default.csv};
\addplot table[ x index = {8}, y index = {9}, col sep=comma] {figures/data/simul_cdf_sbas_default.csv};
\legend{SBAS: 3 nodes, SBAS: 4 nodes, SBAS: 5 nodes, SBAS: 6 nodes, no SBAS}
\end{axis}
\end{tikzpicture}
    }
    \caption{
        Cumulative distribution of SBAS resilience against (randomized) AS-level adversary.
    }
    \label{fig:simul-cdf-dflt-random-adv}
\end{figure}

\subsubsection{SBAS Significantly Enhances Resilience Across Adversary ASes and Customer Locations}
\label{sec:improved-resilience}

We analyze the distribution of prefix-level resilience of routes announced by SBAS against sampled AS-level adversaries performing BGP hijack attacks and compare them to the scenario where SBAS is not used (\cref{fig:simul-cdf-dflt-random-adv}).
%
%
Our results show that SBAS deployment significantly improves routing security across adversary ASes and customer locations even with a small number of BGP announcement nodes. 
Increasing the number of announcement nodes in the SBAS backbone further enhances security. 

\begin{figure}[t]
\centering
\resizebox{0.75\linewidth}{!}{
  \begin{tikzpicture}
\begin{axis}
[
every axis plot/.style={
line width=1.25pt,
},
cycle list/Set1,
cycle multiindex* list={
Set1
    \nextlist
[3 of]linestyles
    \nextlist
},
width = \linewidth, 
height = 0.6\linewidth,
xlabel = {Resilience},
ylabel = CDF,
xlabel style={name=xlabel},
ylabel style={name=ylabel, yshift=-1em},
ymin = 0,
ymax = 1.01,
grid = major,
grid style={dashed, gray, opacity=.5},
legend pos = north west,
legend style = {font = \footnotesize},
]

\addplot table[ x index = {0}, y index = {1}, col sep=comma] {figures/data/simul_cdf_sbas_rpki.csv};
\addplot table[ x index = {2}, y index = {3}, col sep=comma] {figures/data/simul_cdf_sbas_rpki.csv};
\addplot table[ x index = {4}, y index = {5}, col sep=comma] {figures/data/simul_cdf_sbas_rpki.csv};
\addplot table[ x index = {6}, y index = {7}, col sep=comma] {figures/data/simul_cdf_sbas_rpki.csv};
\addplot table[ x index = {8}, y index = {9}, col sep=comma] {figures/data/simul_cdf_sbas_rpki.csv};
\legend{ROV-SBAS: 3 nodes, ROV-SBAS: 4 nodes, ROV-SBAS: 5 nodes, ROV-SBAS: 6 nodes, ROV -- no SBAS}
\end{axis}
\end{tikzpicture}
 }
  \caption{
    Cumulative distribution of SBAS resilience against (randomized) AS-level adversary, assuming adoption of ROV.
  }
  \label{fig:simul-cdf-rpki-random-adv}
\end{figure}
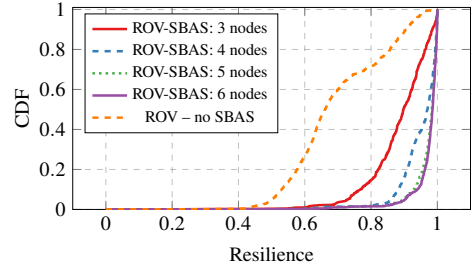

\cref{fig:simul-cdf-dflt-random-adv} shows that 
\textbf{a conservative deployment of SBAS with just 5--6 BGP announcement nodes can lead to more than a 60\% improvement in median resilience.} A three-node SBAS with announcements at Amsterdam, Seattle, and ISI yielded a resilience of $0.750$, a $37.2\%$ improvement compared to the baseline median resilience of $0.545$ (without SBAS). 
Including another announcement node at GRNet further increased median resilience to $0.825$, a $50.9\%$ improvement. Additional announcement nodes provide diminishing security returns: a fifth node at UWashington, to $0.870$ ($59.3\%$ improvement), and finally a sixth node at GATech, to $0.884$ ($61.8\%$ improvement).\\

\paragraph{Resilience improvements against known serial hijacking attack ASes} As an illustrative case study, we examine the threat of BGP attacks launched by a sample of 11 ASes identified as serial hijacking offenders in prior work~\cite{Testart2019}. This smaller sample of attackers allows us to plot resilience for SBAS customers communicating with external hosts for each attacker AS.
Notably, \textbf{SBAS improved resilience against all 11 of the known serial hijacker ASes, with a median resilience improvement of 64.9\%}
 using a configuration of 6 BGP announcement nodes. For the most devastating adversary, AS 9009, the 6-node SBAS produced a median resilience gain of $602.6\%$. (Plotted results and further discussion are included in \cref{app:serial-hijack-simul}.)

\subsubsection{ROV Enforcement in the Broader Internet Can Further Boost Resilience Offered by SBAS} 
We further examine the extent of SBAS's potential security improvements by considering the adoption of ROV (Route Origin Validation using RPKI) 
in the broader (non-SBAS) Internet. 
Recall that ROV makes equally-specific hijacks less likely to succeed by verifying the route's origin AS: an attacker thus needs to prepend itself to a route originating from the valid RPKI-signed AS, which increases the attacker's path length by one. 
%
We repeat the prior simulation setup with the addition of ROV
in the broader internet to evaluate the extent of resilience improvement provided by a combination of SBAS and ROV (see \cref{fig:simul-cdf-rpki-random-adv}).

We use \emph{ROV-aided} to refer to the scenario in which ROV is deployed by the broader internet.
ROV substantially improves resilience offered by SBAS: on average, \textbf{a ROV-aided SBAS deployment increased median resilience over its non-ROV counterpart by 45.0\%.} For example, a ROV-aided SBAS deployment with 3 announcement nodes improves median resilience up to 0.898, a 35.7\% improvement compared to the baseline resilience of 0.661 without SBAS. Similarly, \textbf{a ROV-aided SBAS deployment with 6 announcement nodes improves resilience to 0.985 (a 49.0\% improvement), meaning 98.5\% of adversaries were topologically incapable of hijacking SBAS-announced routes.} 

\vspace{-2mm}
\section{Incentives and Governance}
\vspace{-1mm}
Beyond the setup of the technical SBAS components, adoption of SBAS would require coordination between participating ISPs, formation of organizations to handle governance, and presenting/marketing SBAS to customers. To better understand network operator's SBAS deployment incentives and preference for governance models, we have conducted a survey. 
In Sections~\ref{ref:deployment_incentives} and \ref{ref:governance}, we discuss incentives and possible models of governance for an SBAS deployment, followed by a discussion of survey results in Section~\ref{ref:survey_discussion}.

\vspace{-1mm}
\subsection{Deployment Incentives}
\label{ref:deployment_incentives}

Current market trends demonstrate demand for reliable and secure Internet connectivity. SD-WAN, leased lines, and Network-as-a-service products specifically designed to mitigate routing outages have seen widespread adoption for businesses in various sectors \cite{naas_report}.
The willingness to purchase these services despite substantial costs indicates that customers are willing to pay a premium to protect against routing-induced network outages.  A candidate first customer may have incentives for SBAS's security properties that outweigh the difficulties inevitable in early-stage technology deployment, similar to the initial customer of the SCION network \cite{scion_deployment}. 
The added cost of deploying SBAS is marginal if the infrastructure already supports SCION connectivity~(currently natively supported by 10 ISPs), but the additional customer base that can be reached can provide major financial benefits. Our evaluation in \cref{sec:simulation} has shown that 5 SBAS PoPs can already provide immediate security benefits to the first customer. The current SCION-supporting ISPs would thus suffice for bootstrapping SBAS.



We believe SBAS's 
lightweight implementation will also help it gain early adoption.
Several survey responses~(\cref{ref:survey_discussion}) emphasize the necessity for interoperability with current routing hardware and protocols, with minimum effect on operational robustness as requisite for industry adoption of any new routing security solution. 
Compared to BGP\-sec and other proposed clean-slate routing protocols, SBAS uses commodity network hardware and does not suffer from all-or-nothing deployment security improvements.\footnote{Although the BGPsec signature validation implementation itself can be incrementally deployable, it requires every hop in the routing path to sign its path segments to achieve the desired security properties.}  
SBAS's use of reliable, off-the-shelf
networking components (BGP, iBGP, and secure tunnels) reduces most of the effort required to implement and  maintain custom routing modules. 
This relative ease of implementation translates to lower transition costs for customers and more rapid experimentation under real-world traffic flow conditions. 

\vspace{-1mm}
\subsection{Governance Models}
\label{ref:governance}
Due to the federated nature of the SBAS PoP operation, a governance structure is needed to coordinate global operation~(e.g., AS management, RPKI ROA distribution, and coordination of secure and internal address ranges).
We present four different governance models that all received support in our survey, presented in the order of their degree of centralization and reliance on existing structures.

\textbf{Scenario 1: ICANN and regional Internet registries.}
The regional Internet registries (RIRs) already play a major role in coordinating the control plane of the Internet, e.g., by allocating IP ranges, AS numbers, and providing hosted RPKI services. They would therefore be natural entities to govern a shared AS number for SBAS.
Such a governance model would also benefit from the strong ties between the RIRs and the network providers.
However, albeit they provide coordination activities and services to their members~\cite{ripe_charter, arin_bog_charter,apnic_charter,afrinic_bylaws,lacnic_bylaws}, the RIRs do not cover operation of network infrastructure. Operating SBAS would be orthogonal to other efforts by RIRs to improve routing security.


\textbf{Scenario 2: Multi-stakeholder organization.}
%
Under this governance model, a foundation involving interested parties such as ISPs and companies would run SBAS.
This would provide the benefit of creating an entity with a clear scope of duties with regards to SBAS, entirely dedicated to guarantee the smooth operation of SBAS, and which could also receive dedicated contributions towards that effort.
On the other hand, this would require new structures to be set up.

\textbf{Scenario 3: Federation of network providers.}
%
A governance model relying on the initiative of ISPs working together to join (some of) their resources in SBAS, through a loose coordination at the technical level between the involved parties, building on letters of intent and bilateral agreements.
In this governance model, the network effect is less noticeable and the early participants would bear the bulk of the burden of driving the adoption of the initiative.

\textbf{Scenario 4: Decentralized governance model.}
%
Each PoP operator can join SBAS independently, in the same manner as there is no centralized instance governing which TOR nodes can join the network~\cite{tor_relay_operation}.
This model is most flexible for PoP operators, with low barriers to entry for new operators.
However the continuity of operation of a sufficient number of PoPs is not guaranteed, and sharing scarce resources such as IP address space and AS numbers would be challenging.

\textbf{Governance model recommendation.}
    Based on the results of our survey (presented in Section \ref{ref:survey_discussion}) and the structure of SBAS, we suggest the federation of network providers.
    Not only did this structure receive the most votes in our survey, but it has the benefit of placing the governance decisions in the hands of ISPs, who are the ones responsible for running SBAS PoPs and carrying SBAS traffic.
    Furthermore, it does not require the involvement of any overly powerful or centralized organization, which many operators in our survey expressed concerns over.
    While issues like IP, domain, and ASN allocation inherently involve distribution of a shared resource, governance of SBAS (which operates out of SCION and IP address already controlled by participants) primarily involves technical and policy coordination that can be done in a more decentralized manner.
    We feel that in the same way the MANRS project~\cite{MANRS} has been successful in bringing together ISPs to standardize routing security practices, a federation of ISPs could also be successful in standardizing SBAS operational practices. Additionally, in a real-world operation of SBAS, the federation of providers may establish an association or foundation to create a more concrete structure to govern, organize, and enforce the operation of all participants (incorporating structures of Scenario~2).




\vspace{-1mm}
\subsection{Survey of Network Operators}
\label{ref:survey_discussion}

Survey participants were recruited through direct contact, 
and through the RIPE and NANOG mailing lists \cite{sbas_survey_nanog,sbas_survey_ripe}, garnering 31 responses.
We summarize important results in this section, and discusses detailed survey results in Appendix~\ref{app:survey-details}.

Our survey indicates that network security is very important to the operations of the majority of ISPs and there is a community of early adopters that would be interested in deploying SBAS. Specifically, 26 out of 31 ($84\%$) participants said, on a scale of zero to five, that the importance of network security to their ISP's operations was either a four or a five. However, when asked if secure routing was a marketable product, only 15 operators (of 31) responded with a four or five. 
This gap in responses can be attributed to the usually assumed network effect problem where a large critical mass of participants is needed for substantial security benefits. In these cases, early incentives are insufficient for early stakeholders to undergo the costs of building a new system. This is a major reason why other solutions that require high usages rates to yield security improvements fail to see deployment~(as is the case with BGPsec~\cite{lychev2013bgp}).
Even with substantial interest in the community, many network operators do not see the financial benefits of investing in secure routing, creating a self-fulfilling prophecy of low adoption rates.

We proceeded to ask operators to gauge the interest of their ISP offering SBAS to customers and eight operators rated it as a four or five.  Furthermore, this group of interested operators seemed strongly convinced of the SBAS design: they reported a mean score of 3.75 points on SBAS's deployment feasibility and a mean score of 4.6 on the effectiveness of SBAS against routing attacks. The chief concerns for deployment cited were mostly related to logistics, such as establishing inter-ISP iBGP sessions and several ISPs sharing an ASN in the routing system. While this is a minority of survey participants, it (1) represents 53\% of operators who felt secure routing was a marketable product
and (2) shows there is a non-trivial group of convinced, interested early-adopters that would enable substantial security improvements through SBAS.
%
%
Even with a few early adopters, communication between SBAS participants and the broader Internet achieves substantially higher resilience (e.g., 5 PoPs in \cref{sec:simulation}).

Among the governance models, the federated model was chosen by 14 participants ($\approx$45\%) as the most popular potential governance structure, followed by delegating responsibilities over to the RIRs (35\%), a decentralized model (13\%), and a multi-stakeholder organization (7\%). Several respondents stressed the importance of selecting a structure that would not be dominated by large corporations and with mechanisms to prevent it from growing beyond its needed scope.

By surveying network operators, we see some next steps required for a production deployment of SBAS. We encourage network operators and the research community to work collaboratively to establish SBAS as a production network.



\section{Related Work}
\label{sec:related}

To improve on the limitations of BGP, various alternatives have been suggested, including studies that use overlay technology to establish new routes~\cite{godfrey2009pathlet,gummadi2004improving,yang2006source,apostolaki2019sabre}. Particularly, Andersen~et~al. propose RON (Resilient Overlay Network), an architecture that constructs an overlay network using distributed applications, monitors the underlay routing paths in real-time, and constructs new paths~\cite{andersen2001resilient}. 
Peter~et~al.\ propose the ARROW architecture, which  flattens the Internet topology using overlay tunnels between ISPs, and provides a new route if needed~\cite{peter2014one}. Compared to SBAS, ARROW focuses on availability and only addresses use cases in which customers are fully participating.
%
%
Network pluralism articulates the need for architectural heterogeneity~\cite{turner2005diversifying, koponen2011architecting, bouabene2009autonomic}. 
Crowcroft~et~al. introduce Plutarch, which describes each homogeneous network architecture as context and enables communication across a set of contexts by interstitial functions that interpret the encapsulated functionalities of each context~\cite{crowcroft2003plutarch}. 
Avramopoulos and Rexford present a security backbone framework connecting various secure routing architectures via a secure mesh of virtual links~\cite{avramopoulos2007pluralist}. 
Indeed, network pluralism enables the graceful coexistence of diverse network architectures. 
However, the approaches simply glue network architectures together, supporting them to only \textit{survive}. In contrast, SBAS not only bridges secure routing infrastructures to the Internet in a synergistic manner, but also extends the benefits to the broader Internet, enhancing them to \textit{thrive}.

\section{Conclusion}
\label{sec:conclusion}

While secure routing enjoyed much attention from the research community over the past two decades, real-world adoption has been lagging, perhaps due to the significant infrastructure changes required. With an ambition to make rapid progress to secure routing, we investigate how to leverage a secure communication backbone to secure communication on the regular Internet.
We design and deploy an architecture, SBAS, in which communication between traditional IP endpoints are mediated via a secure backbone that is operated in a federated manner. SBAS substantially reduces the threat of inter-domain routing attacks and only incurs a small latency overhead (and as our results show can even speed up some end-to-end connections compared to the Internet).
A core contribution of this work is the incentive-compatible design. SBAS does not compete with other secure routing architectures, but instead demonstrates that an existing secure routing infrastructure with limited deployment can already benefit the rest of the Internet.
While several challenges still exist when deploying SBAS in a production setting, our survey shows a potential path forward and our experimental results show promise that sizable security improvements can be achieved with even a small set of early adopters. We hope that SBAS revitalizes the quest for secure inter-domain routing.


\section*{Acknowledgements}

This work was supported in part by the National Science Foundation under grants CNS-1553437, CNS-1704105, and CNS-193596 and by the United States Air Force and DARPA under Contract No. FA8750-19-C-0079. Additionally, we gratefully acknowledge support from ETH Zurich, from the ETH4D and EPFL EssentialTech Centre Humanitarian Action Challenge Grant, and from the Zurich Information Security and Privacy Center (ZISC). We would like to thank the anonymous reviewers and our shepherd Deepak Kumar for their valuable feedback as well as David Hausheer, Nicola Rustignoli, Kyveli Mavromati, and the anonymous participants of our survey for their contributions to this project. Any opinions, findings and conclusions or recommendations expressed in this material are those of the authors and do not necessarily reflect the views of the United States Air Force, DARPA, or any other sponsoring agency.

\bibliographystyle{abbrv}
{\small \bibliography{reference}}

\appendix
\section{Overhead of SBAS Components at PoPs}
\label{app:latency}


In comparison to regular packet forwarding, an SBAS needs to perform some additional computational steps, such as decryption from the customer VPN tunnel and encapsulation to remote PoPs.
This first benchmark measures the overhead incurred by the SBAS components on packets passing through a PoP.
To obtain an upper bound on this overhead, we consider communication between a pair of customer hosts.
This is the most computationally intensive scenario, as traffic is sent through VPN tunnels on both sides.

We run ping  from the source customer to the destination host with a $\SI{1}{s}$ interval for an hour.
The overhead is determined by subtracting timestamps of packets captured at the SIG of the PoP from the corresponding timestamps of the same packets captured at the WireGuard interface (after VPN decapsulation, route selection, encapsulation, and tunneling over SCION).
This yields the following results.


\textbf{The majority of packets have sub-millisecond overhead at PoPs}, with a mean of $\SI{0.83}{ms}$ and a standard deviation of $\SI{0.27}{ms}$. The SBAS component delay is also invariant to packet size: repeated experiments with 1KB packets reported a mean overhead of only $\SI{0.75}{ms}$.
We attribute the tail end of the distribution to operating system factors such as process scheduling and resource contention.
Recall that the SBAS components only perform en/decapsulation and routing of packets, exhibiting indistinguishable overhead for different protocols. It is also important to note that, since the current prototype is a software-based implementation, the processing overhead can be further minimized with a production-grade implementation, e.g., using  the Data Plane Development Kit (DPDK)~\cite{dpdk} or hardware accelerators.

\begin{figure}[t]
\centering
\resizebox{0.75\linewidth}{!}{
\begin{tikzpicture}
\begin{axis}
[
every axis plot/.style={
line width=1.25pt,
},
cycle list/Set1,
cycle multiindex* list={
Set1
    \nextlist
[3 of]linestyles
    \nextlist
},
width = 0.9\linewidth, 
height = 0.6\linewidth,
xlabel = {Component overhead [$\si{ms}$]},
ylabel = CDF,
xlabel style={name=xlabel},
ylabel style={name=ylabel, yshift=-1em},
ymin = 0,
xmax = 1.8,
ymax = 1.01,
grid = major,
grid style={dashed, gray, opacity=.5},
legend pos = south east,
legend style = {font = \footnotesize},
]

\addplot table[ x index = {0}, y index = {1}, col sep=comma] {figures/data/latency-overhead-total.csv};
\addlegendentry{Total}

\addplot table[ x index = {0}, y index = {1}, col sep=comma] {figures/data/latency-overhead-vpn.csv};
\addlegendentry[align=center]{
  VPN + Routing \\
 + Encapsulation
}

\end{axis}
\end{tikzpicture}
}
\caption{Overhead incurred by SBAS components at a PoP, measured between the packet arrival from the customer to the sending operation of the SCION packet through SBAS.
        In the dashed data series, the final SCION step is omitted.}
\label{fig:latency-overhead}
\end{figure}
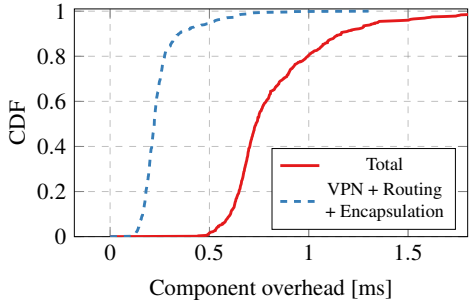

\cref{fig:latency-overhead} summarizes the latency expense incurred by SBAS PoP component in sending a packet through the backbone. 
Moving from an overlay-based network like SCIONLab to a native SCION network would reduce the SBAS overhead further, as the outermost layer of encapsulation (SCION in IP) would not be required anymore.
Moreover, a production-grade implementation of SCION could be used that performs better than the open-source research prototype. The secondary measurement indicated by the dashed line (which omits the SCION latency) in \cref{fig:latency-overhead} provides a lower bound estimate on the potential SBAS backbone latency: approximately $70\%$ of the median $0.74$ ms latency can be attributed to SCION latency, suggesting that the overhead of the SBAS-specific infrastructure is in fact relatively light.

Since our technology is applied to inter-domain traffic, this additional latency is negligible relative to the propagation delay over larger geographical distances, which is often on the order of $\SI{100}{ms}$ and higher.




\ifnum\techreport=1

\section{BGP Route Selection Factors}
\label{app:route-decisions}

The route selection factors of BGP cause a globally distributed infrastructure with many nodes making BGP announcements to generally be more resilient to hijacks, when compared to a case where only a single AS on the Internet makes its BGP announcement (as is the typical case in the absence of SBAS). 
Below is a list of the top BGP route selection factors (in order of preference) with an explanation of how these factors impact routing in the event of a hijack.
\begin{enumerate}
    \item \textbf{Local preference.} A route's local preference is a numerical value attached to a route by an AS when it hears that route in a BGP announcement. Local preference is usually based on which neighbor sent the BGP update: routes from customers (that pay for bytes delivered) are given higher preference over routes from peers (that exchange traffic free of charge), which are further preferred over routes from providers (that must be payed for bytes sent). Given that local preference is one of the first steps in a BGP routing decision, having a large number of transit providers (all of whom will hear your customer route and prefer it over a peer or provider route to the adversary) is beneficial in mitigating the effect of a hijack. SBAS's geographically distributed infrastructure has many different providers and adding additional PoPs will increase the number of providers, strengthening the announcement in the event of a hijack.
    \item \textbf{AS-path length.} After local preference, the length of the AS path in an announcement is used to compare BGP routes with shorter AS-paths being preferred.
    This corresponds loosely to topological proximity in the AS-level Internet graph. With more PoPs, the AS-level distance of a random host to its nearest SBAS PoP will be reduced, thus making the SBAS announcement more preferred.
    \item \textbf{Interior gateway protocol metric.} Presuming a route ties on local preference and AS-path length, the Interior Gateway Protocol (IGP) metric of the next hop router (given in the announcement) is used to break ties. The IGP metric prefers routes that are learned at closer PoPs over more distant ones (from the perspective of the intra-domain routing protocol). Thus, with a geographically-diverse SBAS deployment, the IGP metric of the route to the nearest SBAS PoP is likely to be lower than the IGP metric of a route to a single arbitrary AS.
\end{enumerate}
Thus, by having a global deployment with multiple PoPs, SBAS can offer significantly improved resilience values than a single AS could easily achieve on its own.
\fi

\begin{figure}[t]
    \centering
    \includegraphics[width=\linewidth]{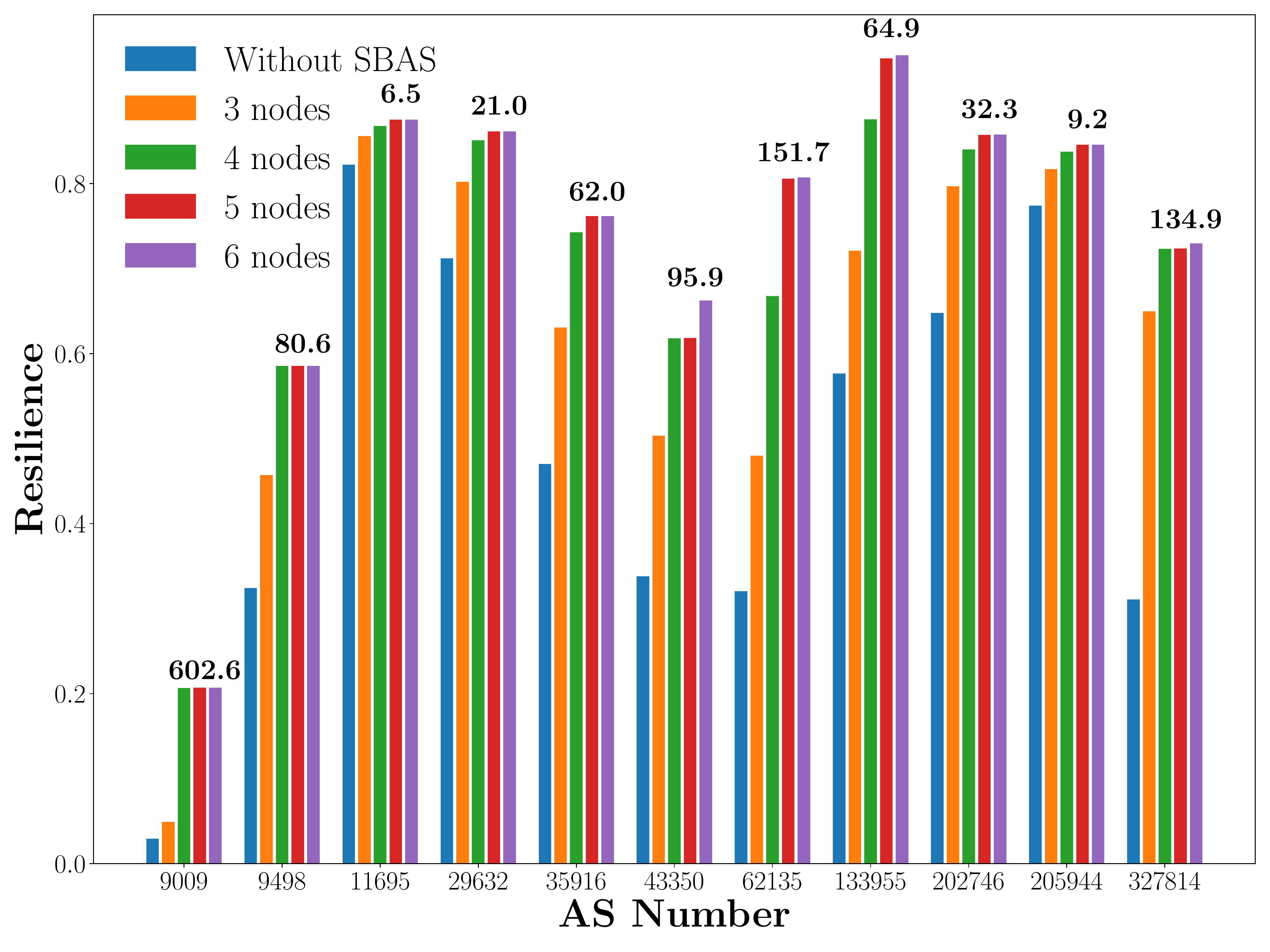}
    \caption{
        Resilience comparison across known serial hijacker ASes. We mark the maximum resilience gain offered by SBAS for each AS adversary. 
    }
    \label{fig:top-adv-barchart}
\end{figure}

\section{Additional Simulation Results}
\textbf{Resilience definition}\label{app:sim:resdef}
For a given victim AS $v \in \mathcal{V}$, we consider a set of potential adversary ASes $\mathcal{A}$ along with a set of potential traffic sources $\mathcal{B}$.
Let us consider an adversary AS $a \in \mathcal{A}$ which attempts to launch an equally specific-prefix hijack attack against a prefix $p$ originated from $v$, and a traffic source $b \in \mathcal{B}$ which sends traffic to $p$.
\begin{gather*}
    \alpha(v, a, b) = 
        \begin{cases}
        1 & \text{if } a \text{ fails to hijack traffic from } b \text{ to } v\\
        0 & \text{otherwise}\\
        \end{cases}
\end{gather*}
In our simulations, selection among equally preferred paths is made via  a random tiebreak.
Aggregating across the adversary and traffic source sets, we compute a normalized resilience measure for the victim: 
\begin{gather*}
    \beta(v, \mathcal{A}, \mathcal{B}) = \sum_{a \in \mathcal{A}}\sum_{b \in \mathcal{B}} \frac{\alpha(v, a, b)}{|\mathcal{A}||\mathcal{B}|}
\end{gather*}

Intuitively, a resilience of 1 indicates that a node in set $\mathcal{B}$ attempting to send traffic to a prefix originated by the victim $v$ will always route its traffic to the true origin, even in the presence of equal prefix length attacks by the attackers in $\mathcal{A}$.

\textbf{Serial Hijacker Simulation}
\label{app:serial-hijack-simul}
The histogram in \cref{fig:top-adv-barchart} shows the results of the simulation of SBAS's resilience against a set of  ASes with a history of serial BGP hijacking attacks (as identified in \cref{sec:improved-resilience}. As previously mentioned, SBAS routes offer higher resilience %
than the baseline approach for all of the 11 serial hijacking ASes, with a mean resilience improvement of $114.7\%$. 
Although this result focuses on a relatively small adversary set, it demonstrates that SBAS provides sizable security improvements against not only randomly sampled adversaries but also notoriously malicious ones.

\textbf{Choice of BGP Announcement Node Locations}
\label{app:node-location}

Locations for the BGP announcement nodes for the three- to six-node SBAS configurations simulated in \cref{sec:improved-resilience} were computed by simulating all possible $N$ node combinations from the set of 15 total transit-carrying PEERING nodes against another random sample of 1000 AS attackers, and then choosing the best performing node locations within them. In general, we found that the globally optimal node location configuration was equivalent to the iterative greedily optimized one, which can serve as an efficient guiding principle for scaling up the SBAS backbone in the future (starting from the current deployment shown in \cref{fig:eval-topology}).

\begin{figure}[tbp]
    \centering
    \includegraphics[width=\linewidth]{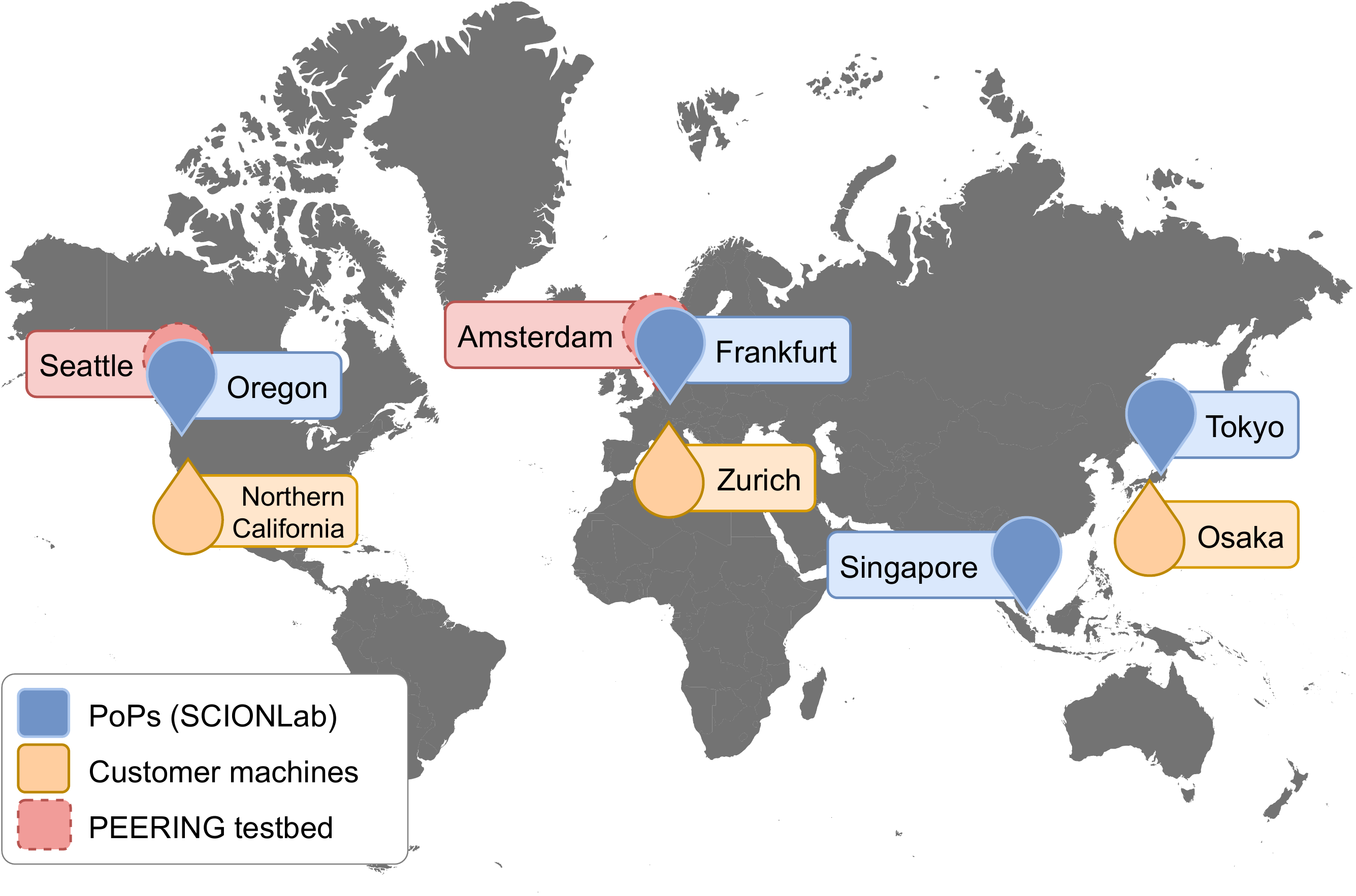}
    \caption{The real-world deployment of SBAS including four PoPs in the United States, Germany, Singapore, and Japan.}
    \label{fig:eval-topology}
\end{figure}

While SBAS may leverage any AS-level participant to announce BGP routes, we restricted the simulation to sets of PEERING nodes as a parallel to deployments in the ethical hijacking experiments.
We note that choosing SBAS BGP announcement nodes outside of the PEERING network may provide further security gains, which will be a focus for future work. We anticipate adding other announcement nodes in geographical areas where PEERING has no presence, such as Asia, may provide the greatest further resilience gains beyond the simulated deployments.

\section{Details of Network Operators Survey}
\label{app:survey-details}


Our survey of network operators consisted of a four-minute video describing SBAS (available at \url{https://youtu.be/xsLjcI-qRd0}) followed by 19 questions divided into four sections: Background, Incentives, Feasibility and Usefulness, and Governance. We designed the survey to not only find out operators opinion's towards SBAS, but also find out what operators thought of secure routing technologies in general since people with different perspectives on secure routing are likely to have different attitudes towards SBAS.
\ifnum\techreport=0
By distributing the survey through direct contact, the RIPE mailing list and the NANOG mailing list, we received 31 responses to the survey. A full description of survey questions and results is contained in our tech report \url{https://arxiv.org/abs/2206.06879}.
\else


Overall we received 31 responses to the survey by distributing the survey through three channels: direct contact with network operators we knew,\footnote{Care was taken such that none of the direct contacts had a close relationship with the research team which could be considered a conflict. Furthermore, all responses were anonymous  and whether a particular contact actually completed the survey is beyond our knowledge.} the RIPE mailing list for network operators, and the NANOG list for network operators. Of the thirty-one responses to the survey, seventeen heard about it through the NANOG mailing list, nine heard about it through direct contact, and two heard about it through the RIPE mailing list. (Two respondents omitted answering the survey source section.)
\fi

\ifnum\techreport=1
\subsection{Background}
As a preamble, the Background section of the survey asks four questions to profile the respondents' experience level, affiliations, and base perspectives towards network security. The questions are listed in the "Background" section of~\cref{tab:survey-questions}.



A summary of the demographics of our survey participants is outlined in Table \ref{tab:survey_demographics}. Our survey captured a sample of network operators with a wide range of experience levels and ISP affiliations. The participants tended to report higher levels of experience, with veteran operators with 20+ years experience constituting more than a third of the participants. The participants also represented a balanced range of different types of ISPs. Operators affiliated with regional ISPs were the most well represented with 12 participants (38.7\%), followed by almost equal numbers at national and research/academic ISPs (each approx. 25\%) and lastly large multi-national ISPs (4 participants, or 13\%). Overall, the survey respondents seemed to value Internet security: more than half of the participants assigned an importance of 5 of 5 points to network security, and only 3 participants reported that security is very low importance or not important at all. 

\begin{table}[t]
    \centering
    \footnotesize
    \begin{tabular}{lcc} 
\textbf{Metric} & \textbf{Percentage} & \textbf{Participants} \\
\toprule
$<$ 1  & 3.2\% & 1 \\
2 - 5  & 19.4\% & 6 \\
6 - 10 & 19.4\% & 6 \\
11 - 20 & 19.4\% & 6 \\
$>$ 20 & 38.7\% & 12 \\
\hline
Direct contact & 25.8\% & 8 \\
RIPE mailing list & 6.5\% & 2 \\
NANOG mailing list & 61.3\% & 19 \\
\textit{Answer not given} & 6.5\% & 2\\
\hline
Large multi-national ISPs & 12.9\% & 4 \\
National ISPs & 25.8\% & 8 \\
Regional ISPs & 38.7\% & 12 \\
Research \& Academic ISPs & 22.6\% & 7 \\ 
\hline
Network security very important 5 & 51.6\% & 16 \\
4 & 32.3\% & 10 \\
3 & 6.5\% & 2 \\
2 & 0\% & 0 \\
1 & 3.2\% & 1 \\
Network security not important at all; N/A & 6.5\% & 2 \\
\bottomrule
\end{tabular}
\vspace{.2cm}
\caption{
        Demographics of SBAS survey participants.
    }
    \label{tab:survey_demographics}
\end{table}

\begin{figure}[t]
\centering
    \includegraphics[width=0.98\linewidth]{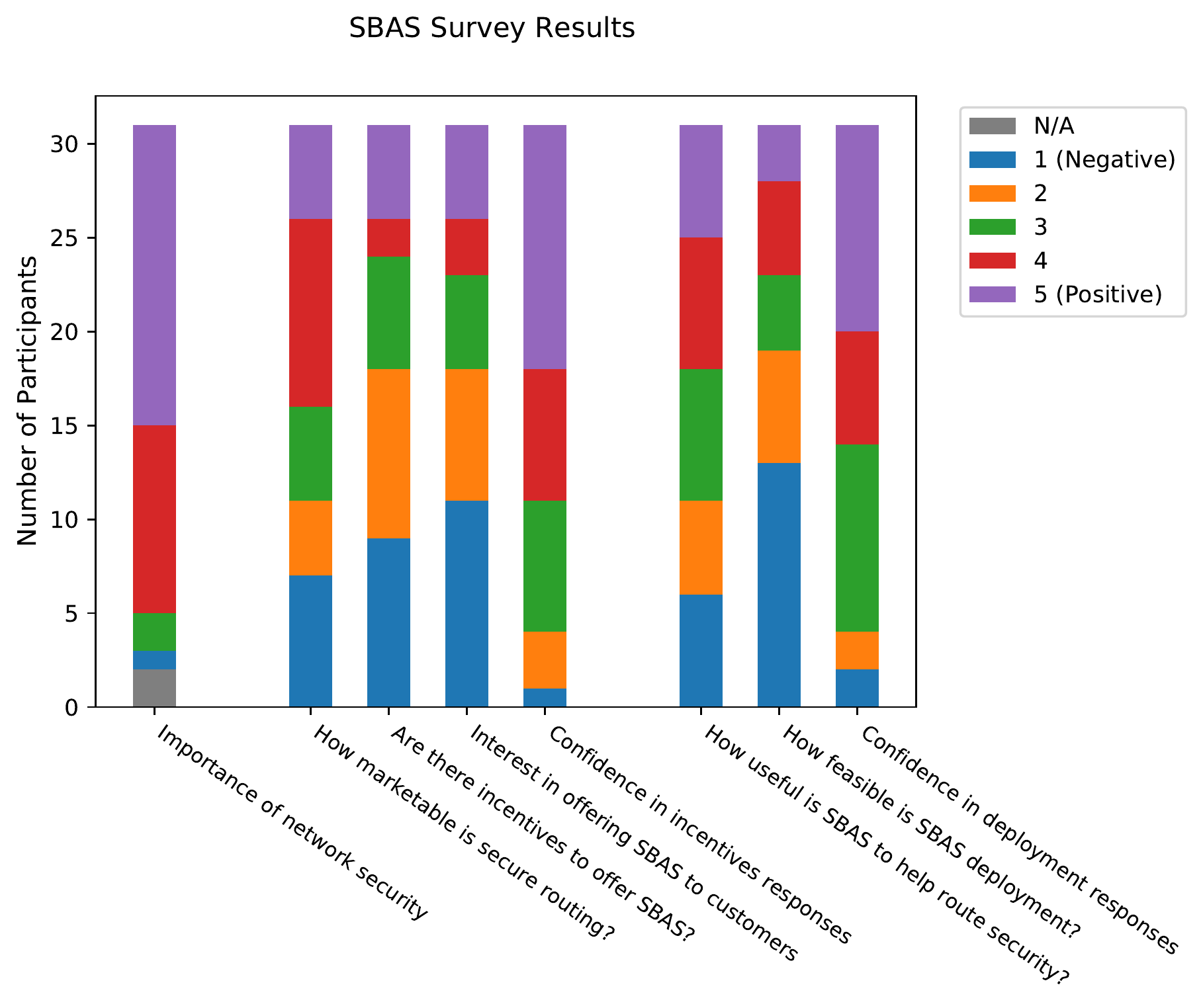}
    \caption{Summarization of survey responses for incentives- and deployment feasibility-related questions.}
    \label{fig:survey-summary}
\end{figure}

\subsection{Incentives}\label{ssec:survey-incentives}
This portion of the survey aimed to capture opinions on incentives for deploying SBAS.
The questions are listed in the "Incentives" section of~\cref{tab:survey-questions}. The breakdown of the survey response to these questions are provided in~\cref{fig:survey-summary}.
Although some operators expressed that they were not interested in deploying SBAS, over 25\% indicated a score of 4 or 5 (very interested) in offering SBAS as a product to customers. 42\% indicated that SBAS presented at least some incentives (a score of 3 points or higher) to network operators to deploy and offer it to customers.


This section also included an optional free-form text response section asking if there were any aspects of SBAS that could help effectively promote it to the management at your ISP. In this response block we got fifteen responses, many of which stated the difficulty of convincing ISP management of the usefulness of such a technology.  Four responses gave insight into specific aspects of SBAS that could help its adoption. These included: 

\begin{itemize}[label={--}]
    \item "Security"
    \item "high-level aspects such as security, ..., ``green''ness"
    \item "Routing selection with help of SION paths. (for customers who have concerns against certain countries or regions.)" 
    \item "Ease of deployment. Cost. Operational overhead."
\end{itemize}
We will keep these points in mind during our future work on this project.

The respondents reported a mean confidence level of 3.9 out of 5 (median 4) in their answers to the incentives questions, indicating they are quite certain in their replies. 

Finally, the section concluded with a text box asking for any other comments regarding incentives. This prompt received four text responses. One response expressed the need for concrete examples of cases of prior attacks where SBAS would have helped. We plan to make this a direction for further research. Another response stated that BGP security was "already here" and an additional "proprietary" solution was not needed.

Another two responses gave us interesting insight into the potential marketing of SBAS. The first stated that "there is no way to sell a "secure" thing when buyers already think the current product is secure. Explaining to them it's not can cost more business than upselling a secure product." This is an interesting point that we will consider when selecting initial participants for SBAS. While some networks are not particularly concerned about the insecurities of current routing infrastructure, there clearly are networks that are (like those that have deployed SCION~\cite{six_scion}) and we plan to focus on these networks as initial adopters.

The second comment expressed doubt toward successfully marketing SBAS to customers: "Usually customer expect a working network at a good price. They are not
 Interested in technical details." We think this is an important point especially as we widen our outreach. We plan to present SBAS in different levels of technical detail tailored to our audience, and construct elegant ways of communicating its security benefits without overwhelming technical details.

\henry{Do we want to interpret the results here in the appendix or simply rely on the interpretation that is in the body?}

\begin{table*}[t]
    \centering
    \footnotesize
    \begin{tabular}{l} 
\textbf{Question} \\
\toprule
\textbf{Background} \\
How many years of experience do you have in network operations? \\ 
--- \emph{1 year or less; 2-5 years; 6-10 years; 11-20 years; 20+ years} \\
Are you currently affiliated with an ISP? \emph{Yes; No} \\
Which group of ISPs best fits the ISP you are affiliated with (could be employed by, if you are currently unaffiliated)? \\ 
--- \emph{Large, multi-national ISPs (Level 3,Telia, Cogent, etc...)} \\  
--- \emph{National ISPs (Comcast, Swisscom, etc...)}; \\ 
--- \emph{Regional ISPs} \\
--- \emph{Research and Academic ISPs (Internet2, GÉANT, NJEdge, GRNet, etc...)} \\
If currently affiliated with an ISP, please rate the importance of network security to your ISP's operations \\ 
--- \emph{0 (not important or not applicable) to 5 (very important)}\\
\hline
\textbf{Incentives} \\
Do you see offering secure routing technology as a marketable product? \\
--- \emph{1 (not at all) to 5 (absolutely)} \\
Would network operators have incentives to offer SBAS as a product to customers? \\
--- \emph{1 (low incentives) to 5 (high incentives)} \\
How interested would you or your ISP be in offering SBAS as a product to customers? \\
--- \emph{1 (not interested) to 5 (very interested)} \\
What is your confidence in your answers to incentives questions? \\
--- \emph{1 (not confident) to 5 (very confident)} \\
\hline
\textbf{Feasibility \& Deployment} \\
How useful is the SBAS system for countering the threat of Internet routing attacks? \\
--- \emph{1 (not useful) to 5 (very useful)} \\
What is the feasibility of deploying SBAS in real-world networks? \\
--- \emph{1 (not feasible) to 5 (very feasible)} \\
What is your confidence in your answers to feasibility and usefulness questions? \\
--- \emph{1 (not confident) to 5 (very confident)} \\
\hline
\textbf{Governance} \\
What governance structure would best encourage ISP participation? \\
--- \emph{Federation of network providers/ISPs (e.g., MANRS)} \\
--- \emph{Multi stakeholder organization involving ISPs, enterprises, developers (e.g., Foundation)} \\
--- \emph{Decentralized governance model (e.g., Tor project)} \\
--- \emph{Handing governance responsibilities over to the Regional Internet Registries (RIRs)} \\
What is your confidence in your answer to the governance question? \\
--- \emph{1 (not confident) to 5 (very confident)} \\
\hline
\bottomrule
\end{tabular}
\vspace{.2cm}
\caption{
        SBAS survey questions. Answer choices are italicized. 
    }
    \label{tab:survey-questions}
\end{table*}

\subsection{Feasibility and Usefulness}\label{ssec:survey-feasibility}
After the incentives section, we asked operators about feasibility and usefulness of SBAS. The questions are listed in the "Feasibility \& Deployment" section of~\cref{tab:survey-questions}. These results are also summarized in~\cref{fig:survey-summary}. 
From the responses, we can see that the operators' reception of SBAS's technical security properties is positive, even while many participants rejected SBAS overall. The general participant group rated SBAS's effectiveness to counter Internet routing attacks with a mean score of 3.1; for respondents who rated routing security as a marketable product, this increased to a mean of 3.8.



This section also included an optional free-form response in which participants could provide further comments regarding feasibility and usefulness. One response raised hesitation saying that ISPs might be reluctant to set up iBGP sessions with other ISPs, and another response brought up concerns over ISPs maintaining their identity in the routing system when sharing an ASN. Both of these responses revolve around the identity of an individual ISP in the SBAS ecosystem and we are thus working on formalizing the route filters and policy boundaries that will exist between ISPs in the SBAS ecosystem.

Another response mentioned that "the extra latency cost limits the usefulness." In similar spirit, we also got a response saying that usefulness was dependent on how SBAS affected reliability and robustness of ISP operations in a worst-case scenario. From these two responses we appreciate the importance of our engineering in ensuring that SBAS is indeed an optimal and highly reliable system that does not substantially degrade ordinary or worse case ISP operations.

\subsection{Governance}\label{ssec:survey-gobernance}
The following section on governance aimed to gauge network operators' preferences with regards to the four proposed operational structures (outlined in~\cref{ref:governance}). The question and its choices are listed in~\cref{tab:survey-questions}. The answers are summarized in~\cref{fig:survey-governance-piechart}. A federation of network providers or ISPs was the most popular option, receiving 14 votes; a multi-stakeholder organization formed by the constituent ISPs and enterprises was the least liked and was selected by only 2 participants. Respondents reported a mean confidence score of 3.7 (median 3) for their answers to this question.


\begin{figure}[tbp]
    \centering
    \includegraphics[width=\linewidth]{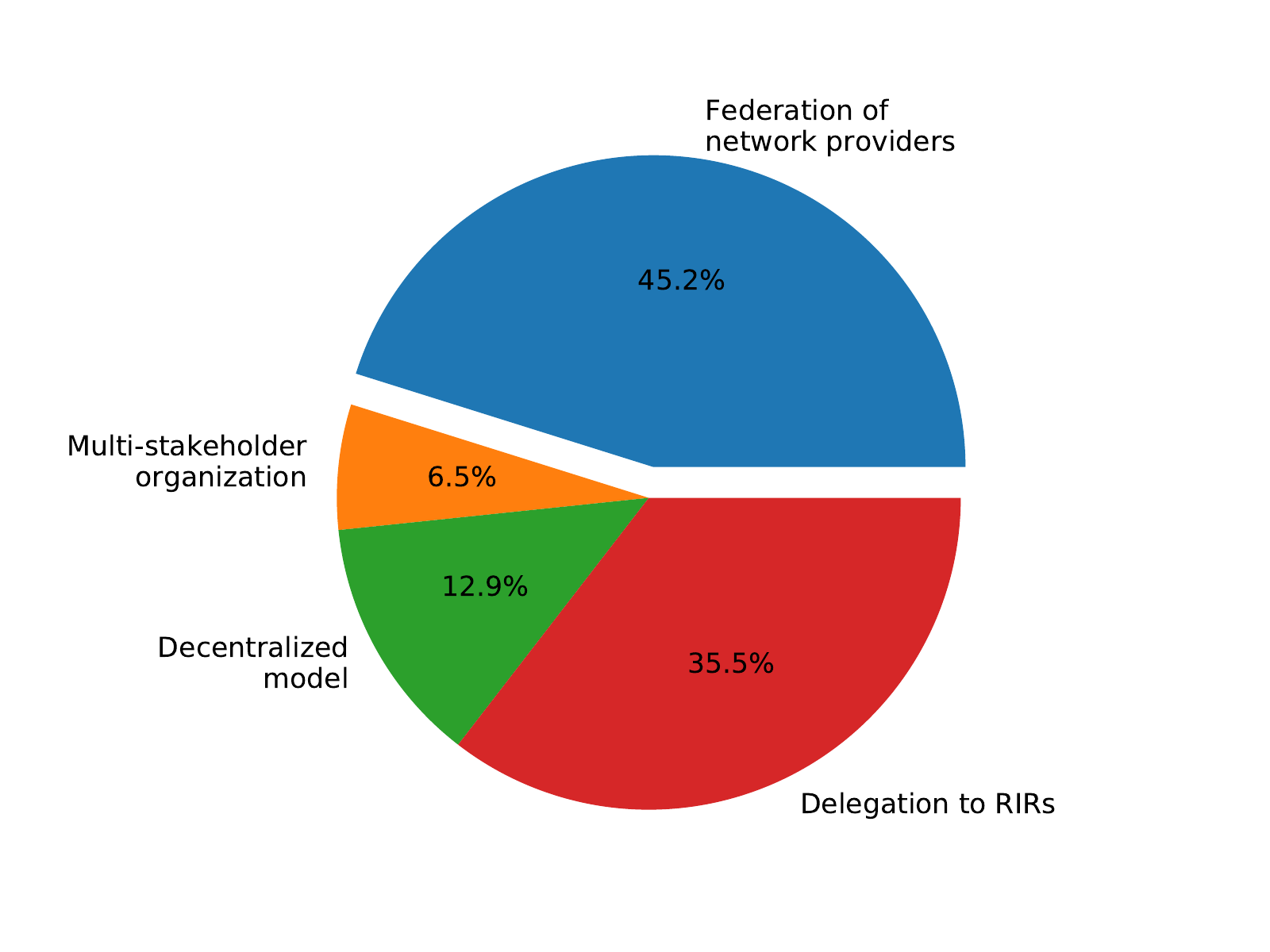}
    \caption{Survey responses for potential SBAS governance models.}
    \label{fig:survey-governance-piechart}
\end{figure}

This section lastly included an optional free-form text question asking if participants had any other comments regarding governance. We got six responses to this question. 

Two responses expressed concerns over a powerful centralized governance structure with one response mentioning that centralized structures can work to the advantage of corporate incumbents and the other citing ICANN as an example of an organization that prioritizes ensuring its existence over fulfilling its mission. We will keep these concerns in mind as we further refine potential governance models.

One response stated that governance should be independent of state-level boundaries to prevent Internet segregation. Another response expressed concern with handing governance over to RIRs with particular concern over ARIN's involvement. We generally feel these comments synergize with the concerns over strong centralized governance.


We ended the survey with an optional free-form text question asking for the respondents' thoughts on the SBAS system overall. To this question we got ten responses.




One response mentioned the benefits are unclear given that prefixes are still announced over the traditional Internet. We also see this as the result of the brevity of our description. As we discuss in our security analysis section, by announcing customer prefixes at multiple PoPs spread throughout the Internet, we can substantially enhance resilience over a standard BGP announcement.

Another response mentioned that the data-plane changes required to devices in the default-free-zone were not viable. We further see this as a result of the brevity of our explanation. The hardware in SBAS customer networks does not need any modification and neither does the hardware in non-participating networks which will still see benefits from SBAS. SBAS PoPs will have to participate in the secure backbone that may require different hardware, but this is already a requirement of upgrading to a secure backbone (independent of SBAS) and recent work has shown potential for running SCION on programmable commodity hardware~\cite{de_ruiter2021scionp4}.

In addition we did get one comment stating the need for a system like SBAS was urgent.

\subsubsection{Summary of Survey Results}
Overall, our survey showed that the opinions of network operators varied tremendously. There is clearly a small group of operators that support and have interest in SBAS. In addition there are operators that are skeptical of SBAS and have a variety of different concerns ranging from incentives and marketability to technical challenges and questions regarding the security benefit. Looking at free-form text responses, we see many technical and security concerns as the result of the brevity of our explanation. Survey length is often inversely proportional to participation, making brevity a necessary trade off. We did link to a draft of our paper for operators that wanted to learn more but we are not aware of any operators that did so (in fact one operator that mentioned confusion said that he or she "should" in fact read our paper).

We intend to take the incentives and marketing concerns very seriously. One insight from our survey is that secure routing technologies are a hard sell: only 45\% of network operators say they list a four or five for whether secure routing is a marketable product while 24\% answered with a score of one implying it was not marketable. Several respondents seemed unconvinced of the threat of hijacking attacks at all: one participant provided the comment that "BGP security is already there." Thus, a crucial challenge in propagating SBAS lies in educating its user base in the importance of improving routing security, and how SBAS can do so well. We plan to focus our adoption on the segment of the community that is interested in deploying such technologies.
Fortunately, providing security benefits to non-participating members when only a fraction of the community participates in SBAS is one of SBAS's primary strengths and we see initially targeting this smaller fraction of the community as a viable strategy for our next steps.
\fi
\section{System Details of an SBAS PoP}\label{sec:appendix:details}

\textbf{Data Plane}\label{sec:appendix:details:data-plane}
In order to handle packets, an SBAS node primarily operates three routing tables for different types of destinations: control, secure, and optimized. When a packet has to egress the kernel of the SBAS PoP (either when coming from the PoP itself or being forwarded), the PoP first checks if the packet contains the source IP of the local router running on the PoP. These are control packets used to enable the iBGP sessions between different SBAS PoPs and are routed using the control table (which contains routes to the IP addresses used by the routers at different SBAS PoPs). This table is loaded with the highest priority and is used exclusively for inter-router communication. 

Next, all packets are checked against the secure table. This table contains the RPKI-validated routes to different SBAS customers. If there is a covering prefix in the secure table, a packet is always routed via this prefix (to the appropriate SBAS customer) \emph{regardless of whether an Internet route for that prefix or a more specific prefix exists}. This prevents routing loops (since customer's secure prefixes are also announced to the Internet) and ensures that, even in the event of a sub-prefix BGP hijack, a packet is sent through the secure network to the right customer.

After this, all packets hit the optimized routing table. This table contains routes to Internet destinations. These routes either involve sending the packet to one of the PoPs Internet peers or providers or sending it to another SBAS PoP in the case where the SBAS PoP (this is appropriate in the case where the SBAS PoP does not make BGP announcements or an alternate PoP's Internet route is selected by the routing engine).

\textbf{Control Plane}
Each SBAS PoP maintains three types of BGP sessions: iBGP sessions with other PoPs, eBGP sessions with SBAS customers, and eBGP sessions with Internet peers and providers. Prefixes learned from each of these sessions are loaded into different tables and handled in the data plane (see Section \ref{sec:appendix:details:data-plane} for more details). Furthermore, for Internet routes and routes learned from other SBAS PoPs, the SBAS routing engine performs route selection between different available routes based on a user configurable metric that can vary from security, to greenness, to preference for certain geographic regions. This is done by having the BIRD routing demon (which manages the BGP sessions) output routes in MRT format~\cite{RFC_MRT} which is then parsed by the SBAS routing engine. The SBAS routing engine then compares the user-defined metric on the available routes in the MRT file and installed the best one into the optimized routing table.

\end{document}